\documentclass[aps,prb,reprint,amsmath,amssymb,notitlepage,citeautoscript,superscriptaddress,showpacs,showkeys]{revtex4-1}

\usepackage{bm,mathrsfs,dcolumn,graphicx,color}
\usepackage{amsmath}
\usepackage{graphicx}
\usepackage[T1]{fontenc}
\usepackage{hyphenat}
\usepackage{setspace}

\usepackage[normalem]{ulem}
\definecolor{darkerblue}{rgb}{0,0,0.75}
\definecolor{darkerred}{rgb}{0.8,0,0}
\usepackage[colorlinks=true,citecolor=darkerblue,linkcolor=darkerred]{hyperref}

\definecolor{ablue}{rgb}{0.1,0.35,0.75}
\definecolor{agreen}{rgb}{0,0.55,0.3}
\definecolor{ared}{rgb}{0.8,0,0}
\definecolor{abrown}{RGB}{160,82,45}
\definecolor{mm}{rgb}{0.5,0.05,0.5}

\usepackage{hyperref}

\usepackage{enumitem} 

\begin{document}

\title{Exciton diffusion in monolayer semiconductors with suppressed disorder}

\author{Jonas Zipfel}
\author{Marvin Kulig}
\affiliation{Department of Physics, University of Regensburg, Regensburg D-93053, Germany}
\author{Ra{\"u}l Perea-Caus{\'i}n}
\author{Samuel Brem}
\affiliation{Department of Physics, Chalmers University of Technology, Fysikg\aa rden 1, 41258 Gothenburg, Sweden}
\author{Jonas D. Ziegler}
\affiliation{Department of Physics, University of Regensburg, Regensburg D-93053, Germany}
\author{Roberto Rosati}
\affiliation{Department of Physics, Chalmers University of Technology, Fysikg\aa rden 1, 41258 Gothenburg, Sweden}
\author{Takashi Taniguchi}
\author{Kenji Watanabe}
\affiliation{National Institute for Materials Science, Tsukuba, Ibaraki 305-004, Japan}
\author{Mikhail M. Glazov}
\affiliation{Ioffe Institute, Saint Petersburg, Russian Federation}
\author{Ermin Malic}
\affiliation{Department of Physics, Chalmers University of Technology, Fysikg\aa rden 1, 41258 Gothenburg, Sweden}
\author{Alexey Chernikov}
\email{alexey.chernikov@ur.de}
\affiliation{Department of Physics, University of Regensburg, Regensburg D-93053, Germany}

\begin{abstract}
Tightly bound excitons in monolayer semiconductors represent a versatile platform to study two-dimensional propagation of neutral quasiparticles.
Their intrinsic properties, however, can be severely obscured by spatial energy fluctuations due to a high sensitivity to the immediate environment.
Here, we take advantage of the encapsulation of individual layers in hexagonal boron nitride to strongly suppress environmental disorder.
Diffusion of excitons is then directly monitored using time- and spatially-resolved emission microscopy at ambient conditions.
We consistently find very efficient propagation with linear diffusion coefficients up to 10\,cm$^2$/s, corresponding to room temperature effective mobilities as high as 400\,cm$^2$/Vs as well as a correlation between rapid diffusion and short population lifetime. 
At elevated densities we detect distinct signatures of many-particle interactions and consequences of strongly suppressed Auger-like exciton-exciton annihilation.
A combination of analytical and numerical theoretical approaches is employed to provide pathways towards comprehensive understanding of the observed linear and non-linear propagation phenomena.
We emphasize the role of dark exciton states and present a mechanism for diffusion facilitated by free electron hole plasma from entropy-ionized excitons.
\end{abstract}
\keywords{Two-dimensional materials, Transition-metal dichalcogenides, Excitons, Diffusion}
\maketitle

\section{Introduction}

A key property of low-dimensional semiconductors with spatial degrees of freedom is their inherent capability to facilitate transport of photoexcitations\,\cite{ivchenko05a}. 
In addition, effects of quantum confinement and reduced dielectric screening can result in exceptionally strong Coulomb interactions\,\cite{Rytova1967, Keldysh1979, Klingshirn2007, Haug2009}.
As a consequence, propagation of photoexcited charge carriers becomes highly correlated as electrons and holes form bound exciton states\,\cite{Frenkel1931, gross:exciton:eng}. 
The excitons can dominate the response to external fields and lead to dramatic changes in the transport properties of the optical excitations with immediate fundamental and technological implications.

Prominent representatives of low-dimensional systems combining translational symmetry and efficient Coulomb coupling are semiconducting transition metal dichalcogenides (TMDCs)\,\cite{Wilson1969}.
In their bulk form, TMDCs are van der Waals crystals composed of stacked $MX_2$ layers, typically, with $M$ = Mo, W and $X$ = S, Se, Te.
On the nanoscale, they can form stable structures down to the effective two-dimensional limit of single monolayers as thin as a few angstroms\,\cite{Novoselov2005, Mak2010, Splendiani2010}.
The TMDC family was shown to host a rich variety of remarkable properties associated with their peculiar electronic structure, spin-valley locking, strong light-matter interactions, and complex many-body physics\,\cite{Xu2014,Yu2015,Xiao2017,Wang2018}, stimulating broad interest across the research community.

More recently, an increasing amount of efforts was directed to explore spatial properties of photoexcitations in TMDC monolayers\,\cite{Kumar2014,Mouri2014,Kato2016,Yuan2017,Cadiz2018,Kulig2018,CordovillaLeon2018,Wang2019,Glazov2019,Shahnazaryan2019,Hao2019,Perea-Causin2019,Fu2019,CordovillaLeon2019} in view of robust exciton states combined with their freedom to move in two dimensions.
The excitons were found to be mobile with diffusion lengths up to many 100's of nanometers at ambient conditions\,\cite{Kumar2014,Kato2016,Yuan2017,Cadiz2018,Fu2019}, guided by strain and dielectric gradients\,\cite{CordovillaLeon2018,Shahnazaryan2019,Hao2019}, as well as exhibiting intriguing non-linear phenomena during spatial propagation associated with efficient interactions\,\cite{Mouri2014,Kulig2018,Wang2019,Glazov2019,Perea-Causin2019}.
In heterostructures, the complex nature of the exciton transport was further highlighted by the demonstration of non-trivial diffusion of spin-valley polarized excitations\,\cite{Rivera2016}, interlayer exciton diffusion\,\cite{Calman2018}, and external control of the exciton currents\,\cite{Unuchek2018}.
Two-dimensional TMDCs thus emerged as a highly promising solid-state platform for experimental and theoretical inquiries into the physics of quasiparticle propagation in complex multi-component systems on the nanoscale.

However, ultra-thin materials such as TMDCs are particularly susceptible to environmental disorder\,\cite{Raja2019}, strongly affecting both optical and transport properties.
Following the original work on graphene\,\cite{Dean2010}, encapsulation of individual layers of TMDCs in hexagonal boron nitride (hBN) recently emerged as a key technology to mitigate inhomogeneities introduced by the materials' surroundings\,\cite{Cadiz2017, Ajayi2017, Courtade2017, Manca2017, Stier2018}.
It provided direct access to a number of inherent properties of TMDCs including radiative broadening of optical transitions, reliable identification of exciton complexes and their behavior, as well as the demonstration of high electronic mobilities.
These recent advances strongly motivate to explore linear and non-linear exciton propagation in light of the rapid progress in material preparation, as exemplified by the initial reports of efficient diffusion in hBN-encapsulated WS$_2$ and WSe$_2$ monolayer samples\,\cite{Cadiz2018,Fu2019}.
The improved material platform should provide unique opportunities to access the physics of mobile photoexcitations in monolayer semiconductors in essentially disorder-free environments.

Here, we present the results of a joint experimental and theoretical study of exciton diffusion in hBN-encapsulated WS$_2$ monolayer samples.
Using spatially- and time-resolved micro-photoluminescence we directly monitor spatial propagation of photoexcited electron-hole pairs after optical injection at ambient conditions.
In the linear regime, we consistently find exceptionally rapid diffusion with coefficients up to 10\,cm$^2$/s, corresponding to room temperature effective mobilities as high as 400\,cm$^2$/Vs.
At increased densities, we observe pronounced non-linearities in the propagation due to many-particle interactions, but also the consequences of strongly suppressed exciton-exciton annihilation.

To provide comprehensive, quantitative understanding of our findings, we employ a combination of analytical and numerical theoretical approaches to address both linear and non-linear regimes of exciton diffusion.
We discuss the influence of a complex exciton bandstructure
and emphasize the role of both bright and dark states of the excitons that are important to understand the spatial properties of the studied system. 
In addition, we discuss the consequences of the entropy ionization of excitons that can lead to large fractions of free carrier populations and determine the diffusion in the low excitation density regime.
For elevated densities we demonstrate the impact of reduced Auger-like recombination for the propagation of optical excitations considering both enhanced broadening of the spatial distribution and halo-like phenomena.

The manuscript is organized as follows. 
In Sec.\,\ref{exp} we outline the details of the material fabrication and experimental procedures. 
Section \ref{results} provides an overview of the main experimental observations with an illustration of the key properties of density-dependent exciton propagation in hBN-encapsulated TMDCs.
The analysis of the low-density regime is given in Sec. \ref{linear}, focused on the exciton diffusion in multi-component systems and the interplay with the non-radiative recombination dynamics, including quantitative theoretical description of the studied processes.
Non-linear phenomena related to Auger-like recombination are then discussed in Sec. \ref{non-linear}, illustrating the consequences of suppressed exciton-exciton interaction through analytical and numerical models.
The main findings and conclusions are summarized in Sec. \ref{conclusions}.

\section{Experimental details}
\label{exp}
 
The studied samples were obtained by mechanical exfoliation of chemically synthesized bulk crystals (WS$_2$ from ``HQgraphene'', hBN from NIMS) onto polydimethylsiloxane films. 
Subsequently, individual flakes were stamped following the procedure from Ref.\,\onlinecite{Castellanos-Gomez2014a} onto 70$^o$C pre-heated SiO$_2$/Si substrates starting with a thin hBN layer, followed by a WS$_2$ monolayer that is then capped by a thin hBN layer.
The average thickness of the hBN flakes estimated from optical contrast and supported by atomic-force-microscopy was on the order of 5 to 15\,nm.
The fabrication method typically provided hBN/WS$_2$/hBN heterostructures with an average area of several 100's of $\mu$m$^2$.

The samples were first analyzed by linear reflectance and continuous-wave photoluminescence mapping at cryogenic and room temperature.
All measurements were performed on freshly-exfoliated samples that showed consistent emission at the fundamental neutral exciton resonance (Appendix \ref{AppC} illustrates changes in the luminescence spectra detected one year after the initial measurements).
For the subsequent experiments, sufficiently large regions in the range of 10's to 100 $\mu$m$^2$ with good interlayer contact between TMDC monolayers and the surrounding hBN flakes were identified.
Key prerequisites were characteristically narrow spectral linewidths of the exciton ground and excited state resonances, limited to several meV at low temperatures\,\cite{Selig2016, Brem2019}.
These regions were then used for optical measurements of exciton diffusion, as schematically illustrated in Fig.\ref{fig1}\,(a).
A short laser pulse from the 100\,fs Ti:sapphire laser operating at the repetition rate of 80\,MHz was tightly focused on the sample by a 100x microscope objective.
The resulting full-width at half-maximum of the laser spot was 0.6\,$\mu$m.
The photon energy of the incident beam was tuned to 2.4\,eV, roughly corresponding to a non-resonant excitation of the WS$_2$ monolayer into the flank of the spin-orbit split B-exciton\,\cite{Li2014}.

All measurements were performed at room temperature and ambient conditions.
In this regime, the excitons form rapidly in TMDC monolayers on sub-picosecond timescales\,\cite{Ceballos2016,Steinleitner2017}, thermalize to the lattice temperature within a few picoseconds\,\cite{Selig2018}, diffuse and finally recombine on the timescale of 10's to 100's of picoseconds in typical samples.
Due to the comparatively fast formation and thermalization, the radiative recombination of the small population of bright excitons within the radiative cone allows us to follow the dynamics of the total exciton population, as discussed in Ref.\,\onlinecite{Rosati2019}.
Spatially- and time-resolved emission intensity $I_{PL}(r,t)$ is then monitored by imaging the luminescence cross-section onto a streak-camera detector (see Ref.\,\onlinecite{Kulig2018} for additional details).
Using this technique we directly obtain time-dependent profiles of the spatially-resolved exciton emission with the temporal resolution of a few picoseconds.

\section{Exciton diffusion}
\label{results}

\subsection{Linear regime}

\begin{figure}[t]
	\centering
			\includegraphics[width=8.2 cm]{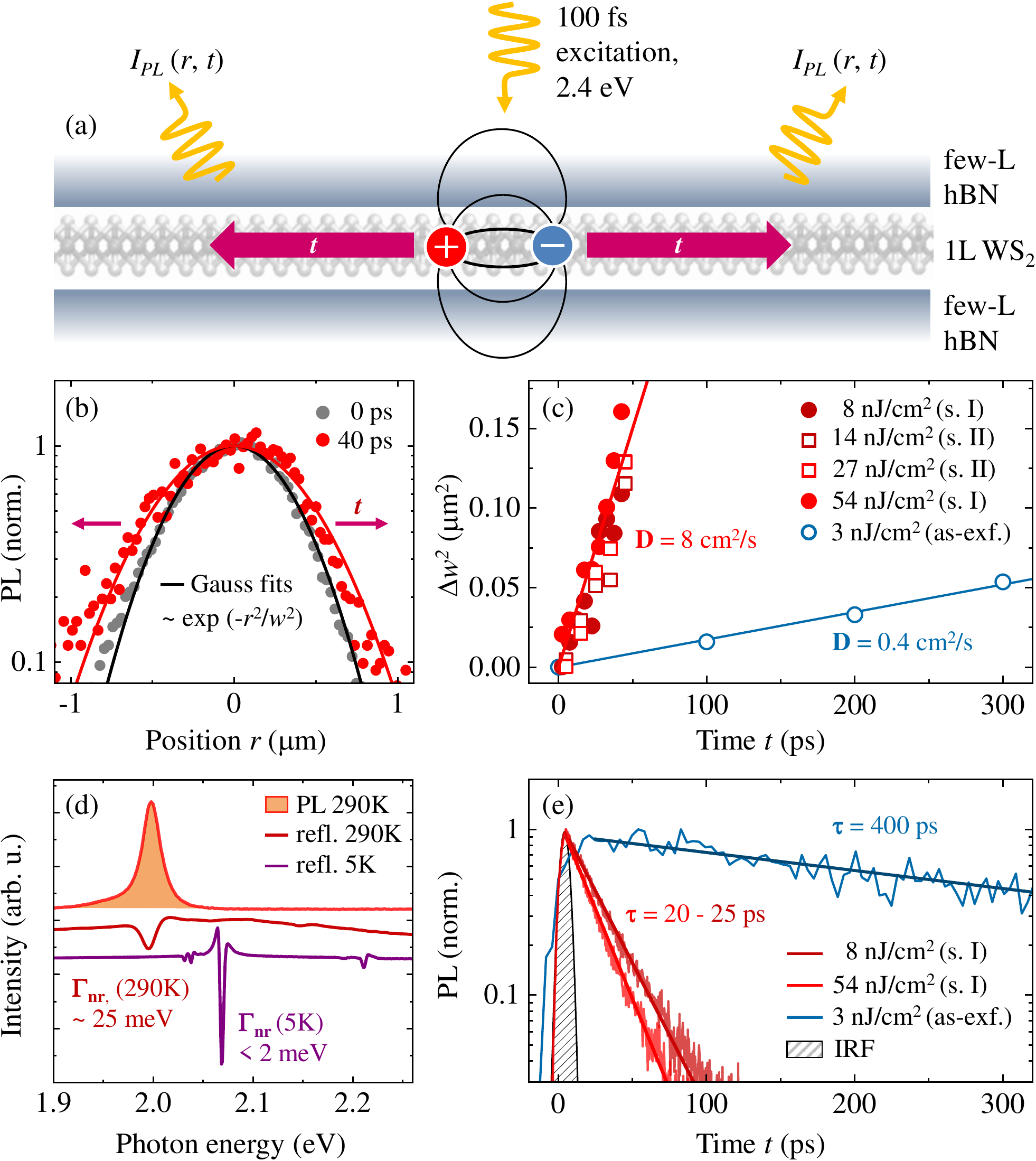}
		\caption{(a) Schematic illustration of the spatially- and time-resolved photoluminescence microscopy on hBN-encapsulated WS$_2$ monolayer samples. 
		(b) Representative emission cross-sections immediately after the excitation at 0 ps and for 40 ps. 
		(c) Extracted relative change of the squared width of the luminescence profiles as function of time after the optical injection. 
		The data is shown for several pump energy densities in the low-density regime, for two representative hBN-encapsulated flakes (I and II) and for a bare, as-exfoliated monolayer on SiO$_2$/Si substrate (taken from Ref.\,\onlinecite{Kulig2018}). 
		(d) Corresponding linear spectroscopy data confirming narrow spectral lines resulting from hBN-encapsulation. 
		(e) Time-resolved PL transients from the hBN-encapsulated samples in direct comparison with an as-exfoliated flake.
		Instrument response of the experimental setup is denoted by IRF.
		}
	\label{fig1}
\end{figure}

Typical spatial luminescence profiles of a hBN-encapsulated WS$_2$ monolayer are presented in Fig.\,\ref{fig1}\,(b) on a semi-logarithmic plot immediately after the excitation at 0\,ps and at a later time, for 40\,ps.
The data already indicates rapid, time-dependent spatial broadening of the exciton distribution.
For quantitative analysis, spatially-resolved PL intensity profiles $I_{PL}(r,t)$ are fitted at each time step $t$ by a Gaussian function, $\propto\exp{[-r^2/w^2(t)]}$.
The squared width $w^2(t)$ of the spatial distribution can be then separated in two components, according to $w^2(t) = w_0^2 + \Delta w^2(t)$: the time-\textit{independent} broadening $w_0^2$ determined by the initial PL profile at $t$=0 convoluted with the instrument point spread function, and the time-\textit{dependent} contribution $\Delta w^2(t)$. 
We note that due to the squared sum rule for the widths of two convoluted Gaussians, the $\Delta w^2(t)$ contribution is not expected to depend on the initial spot width and broadening from the instrument response. 

Figure \,\ref{fig1}\,(c) shows time-dependent traces of the increase of squared emission width obtained from two representative samples of hBN-encapsulated WS$_2$ monolayers in the linear excitation regime.
For comparison we also include the data from a typical as-exfoliated WS$_2$ flake on SiO$_2$/Si substrate, taken from our previous study\,\cite{Kulig2018}.
An effective diffusion coefficient $D$ is then extracted from the slope following the solution of the linear diffusion equation, according to $\Delta w^2(t) = 4Dt$.
For the presented data at low injection densities, $\Delta w^2(t)$ increases linearly with time and does not depend on the pump power.
We note, however, that $D$ can be generally both density- and time-dependent if the propagation dynamics deviate from the linear diffusion law.
In addition, extraction of the diffusion coefficient can be further influenced by density-dependent recombination processes, as previously demonstrated for bulk Cu$_2$O\,\cite{Warren2000} and monolayer TMDCs\,\cite{Kulig2018}, discussed in Sec. \,\ref{non-linear}.

Interestingly, as illustrated in Fig.\,\ref{fig1}\,(c), the slope of $\Delta w^2(t)$ is roughly 20 times steeper for the hBN-encapsulated monolayers in contrast to the as-exfoliated one.
That corresponds to an increase of the diffusion coefficient from 0.4\,cm$^2$/s to 8\,cm$^2$/s. 
Such efficient exciton propagation in the hBN-encapsulated samples is accompanied by characteristic spectral properties, presented in Fig.\,\ref{fig1}\,(d).
One of the most important consequences of encapsulation is exemplified by narrow spectral widths of the resonances.
In general, at elevated temperatures, including the room temperature where our diffusion measurements are performed, the main broadening mechanism of the optical transition stems from the exciton-phonon scattering due to the thermal activation of the phonon modes~\cite{Selig2016, Brem2019}.
For the studied room temperature conditions, the linewidths are thus usually dominated by the homogeneous component with additional, weaker contributions from the phonon-assisted sidebands at the low-energy side of the emission spectrum\,\cite{Christiansen2017}.
In contrast, cryogenic temperatures strongly suppress the exciton-phonon interaction by reducing the phase space for the optical and intervalley acoustic phonon emission and decreasing the number of available low-energy acoustic phonons. 
Low-temperature conditions thus allow for a better analysis of the inhomogeneous contributions from disorder.

Measured room temperature spectra from reflectance and continuous-wave PL shown in Fig.\,\ref{fig1}\,(d) already exhibit exciton linewidths on the order of 25\,meV that correspond reasonably well to the theoretical predictions for intrinsic scattering in monolayer WS$_2$\,\cite{Selig2016, Raja2018}. 
Most importantly, though, both exciton ground and excited state resonances are extremely narrow in the 5\,K reflectance derivative data in Fig.\,\ref{fig1}\,(d), as it is typically observed for successfully encapsulated TMDCs\,\cite{Cadiz2017, Ajayi2017, Courtade2017, Manca2017, Stier2018}. 
In addition, the optical response is dominated by the neutral exciton resonance with only weak signatures at lower energies related to the presence of free carriers, indicating small residual doping\,\cite{Wang2018}.
\begin{figure*}[t]
	\centering
			\includegraphics[width=13.5 cm]{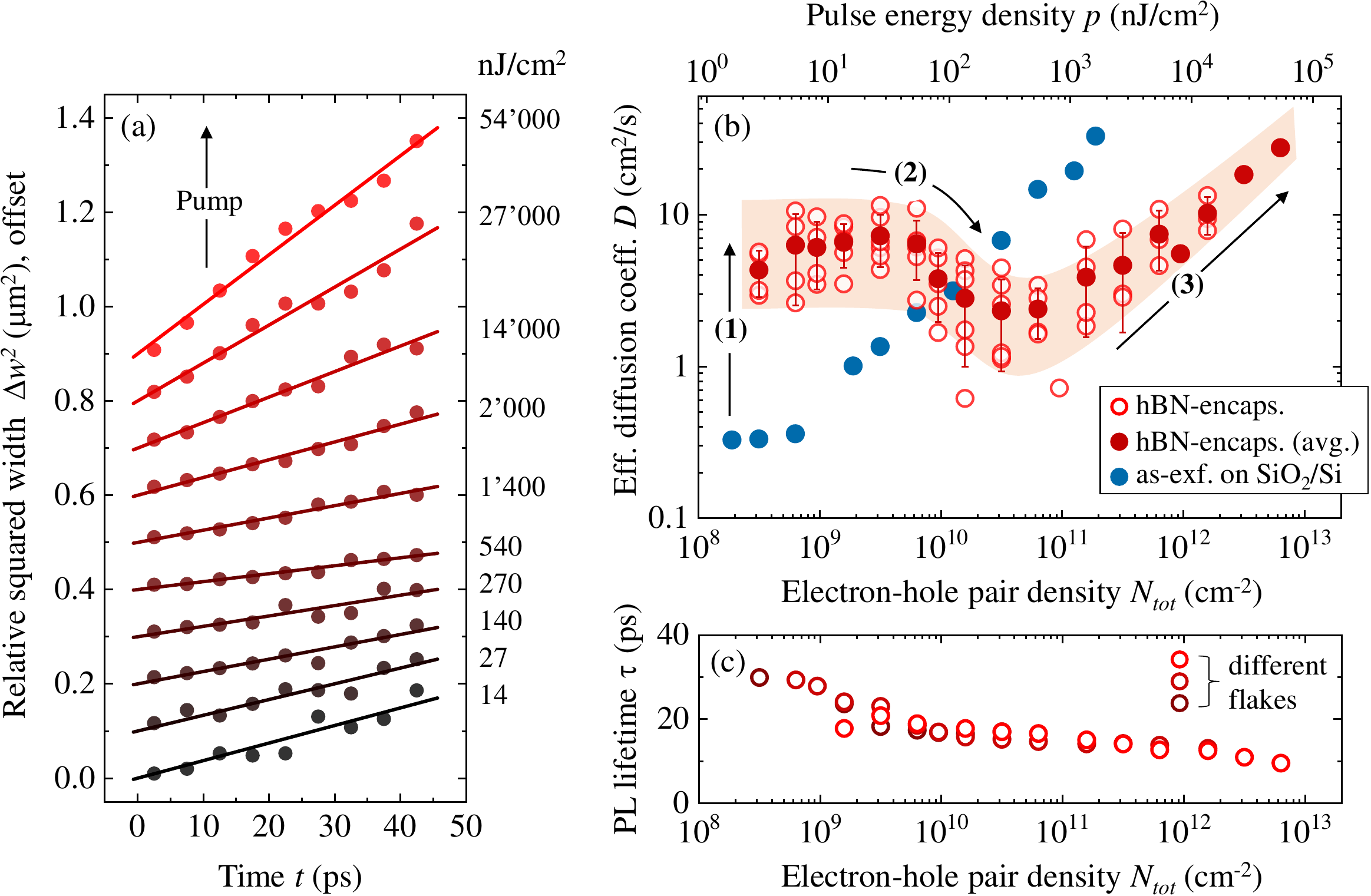}
		\caption{(a) Relative change of the squared width of the luminescence profiles as function of time after the excitation presented for a number of pump energy densities from $1.4 \times 10^1$ to $5.4 \times 10^4$ nJ/cm$^2$.		
		The data is vertically offset for better comparison. 
		Effective diffusion coefficients are extracted from the slopes of the linear fits, shown as solid lines. 
		(b) Overview of the extracted effective diffusion coefficients from several individual measurements (open circles) on hBN-encapsulated WS$_2$ samples together with the averaged result (solid red circles) as function of injected electron-hole pair density. 
		The shaded area roughly illustrates the general density dependence of the measured values.
		Effective diffusion coefficients from an as-exfoliated sample on SiO$_2$/Si substrate are added for comparison, taken from Ref.\,\onlinecite{Kulig2018}. 
		They are plotted for the same electron-hole pair energy densities taking into account about factor of two higher effective absorption of the as-exfoliated sample (the top abscissa axis should be thus understood as being shifted by factor of two to higher densities in this case, c.f. Ref.\,\onlinecite{Kulig2018}).
		(c) Corresponding PL decay times of the hBN-encapsulated samples for several representative flakes.
		}
	\label{fig2}
\end{figure*} 

A key prerequisite for the narrow linewidths is a homogeneous dielectric environment of the monolayer provided by the encapsulation between atomically flat hBN flakes.
The procedure essentially removes dielectric disorder that otherwise arises from local fluctuations of the external dielectric permittivity and can dominate inhomogeneous broadening in as-exfoliated flakes\,\cite{Raja2019}.
The absolute value of the non-radiative linewidth of the exciton ground state extracted from the reflectance spectrum in Fig.\,\ref{fig1}\,(d) is below 2\,meV, already limiting potential contributions from disorder to this very small energy range.
However, the actual residual inhomogeneous contribution in our samples is likely to be even less than 1\,meV, since we already expect an intrinsic non-radiative broadening in WS$_2$ of a few meV due to the relaxation of the $K$-point excitons towards dark intervalley states\,\cite{Selig2016, Raja2018, Brem2019}. 
The impact of disorder should be thus altogether negligible at elevated temperatures, so that the hBN-encapsulated monolayers can be considered as essentially disorder-free for the room-temperature propagation studies.

In this context we emphasize that the observed mitigation of the spatial fluctuations in the average potential, i.e., long-range \textit{disorder}, does not generally imply complete absence of \textit{all imperfections}.
For example, the presence of deep trap states is not expected to strongly contribute to the linewidth broadening, that is dominated by intrinsic ultra-fast exciton-phonon scattering at elevated temperatures and by radiative recombination at liquid helium temperature, limiting the exciton \textit{polarization} lifetime (dephasing time)\,\cite{Selig2016,Brem2019}. 
Instead, propagation of the photoexcited carriers towards the traps and the subsequent capture strongly influences the total \textit{population lifetime} of the photoexcited electron-hole pairs and the overall quantum yield of the material\,\cite{Amani2015}.
In contrast to the phonon-induced polarization dephasing, these processes typically occur on much longer time-scales from a few up to 100's of picoseconds.
As a consequence, the decay of the PL transients is usually considered to be limited by such non-radiative recombination mechanisms, in particular, at higher temperatures.

Interestingly the PL lifetime of the studied samples seems to consistently \textit{decrease} upon hBN-encapsulation, from many 100's to a few 10's of picoseconds, as illustrated by representative transients in Fig.\,\ref{fig1}\,(e).
Moreover, the observed change of the characteristic decay constants is roughly proportional to the difference of the diffusion coefficients. 
A possible mechanism for this behavior originating in the interplay between propagation and non-radiative capture is discussed in detail in Sec.\,\ref{linear}.

\subsection {Density dependence}

To study the influence of interactions between photoexcited electron-hole pairs, exciton diffusion experiments were performed across a broad range of pump densities, from a few nJ/cm$^2$ to 10's of $\mu$J/cm$^2$.
Representative time-dependent squared widths $\Delta w^2 (t)$ of the spatial emission profiles are shown in Fig.\,\ref{fig2}\,(a).
Overall, the exciton expansion $\Delta w^2(t)$ is found to change both strongly and non-monotonously between low and high density regimes.
From the linear fits we extract density-dependent effective diffusion coefficients during the time interval of 0\,\ldots\,40\,ps after the excitation, roughly corresponding to the lifetime of more than 90\,\% of the injected population.
The corresponding individual values are presented in Fig.\,\ref{fig2}\,(b) for varying flakes and measurement positions together with the arithmetic average.
The results from an as-exfoliated sample on SiO$_2$/Si substrate are added for direct comparison.
The upper abscissa axis indicates the average energy density per pump pulse within the width of the excitation spot and the lower abscissa axis corresponds to the estimated injected total electron-hole density $N_{tot}$.
For the latter we used the absorption coefficient of 5\,\% at the chosen pump energy, as obtained from the linear reflectance measurements combined with the analysis of the dielectric function and local field effects from multi-layer interference in hBN-encapsulated samples.
We note that all excitation densities (up to several 10$^{12}$\,cm$^{-2}$) used in our experiments are estimated to be well below the Mott threshold\,\cite{Haug2009,Snoke2008} of exciton ionization that should be close to 10$^{13}$\,cm$^{-2}$ or above\,\cite{Chernikov2015c,Steinhoff2017}.

For the exciton propagation, three characteristic regimes are identified, roughly indicated by (1), (2), and (3) in Fig.\,\ref{fig2}\,(b).
At electron-hole pair densities below 10$^{10}$\,cm$^{-2}$ we observe a constant diffusion coefficient in the range between 5 and 10\,cm$^2$/s within experimental spread.
These values are much higher than in the as-exfoliated samples and rather close to those recently reported for a hBN-encapsulated WSe$_2$ monolayer\,\cite{Cadiz2018}.
In the intermediate density regime between 10$^{10}$ and 10$^{11}$\,cm$^{-2}$ the diffusion coefficient first decreases down to values of about 1 to 3\,cm$^2$/s. 
Finally, it increases again for densities above 10$^{11}$\,cm$^{-2}$ and reaches up to 30\,cm$^2$/s at 6\,$\times$\,10$^{11}$\,cm$^{-2}$.
Interestingly, the population lifetime exhibits only a weak density dependence across the studied excitation range, as illustrated by PL decay times presented in Fig.\,\ref{fig2}\,(c), in good agreement with a previous report\,\cite{Hoshi2017}.
Here, we note that the density-dependent increase of the effective diffusion coefficient is also observed in as-exfoliated WS$_2$ samples, albeit at much lower excitation densities (blue points in Fig.\,\ref{fig2}\,(b)).
It is a common observation for systems exhibiting superlinear density dependent recombination channels\,\cite{Warren2000, Kulig2018, Deng2020}, as discussed in detail Sec.\ref{non-linear}.

In the following we focus on the individual aspects of these experimental observations in the exciton propagation, discuss the underlying mechanisms, and relate our findings to the properties of the interacting many-particle system of electron-hole states in the studied monolayers.

\section{Linear regime analysis}
\label{linear}

In this section we discuss the exciton propagation in the low density regime, where non-linearities are unexpected, and for densities where the first transition towards a non-linear behavior occurs.
These regions are marked by (1) and (2) in Fig.\,\ref{fig2}\,(b), respectively.
First, we focus on the mechanisms determining the diffusion of excitons within a complex multi-valley bandstructure and illustrate the influence of dark states and the electron-hole plasma on the spatial properties of the photoexcited carrier system.
Subsequently we discuss the observation of short population lifetimes in hBN-encapsulated samples that accompanies rapid diffusion.
We outline potential mechanisms of this relationship and illustrate a model scenario for the interplay of propagation and non-radiative capture.

\subsection {Multi-valley exciton diffusion}
\label{linear-A}

First, we address a basic question: \textit{What determines room-temperature exciton diffusion in the studied TMDC monolayers}?
Immediately, one could consider the total scattering time $\tau_s$ of about 30\,fs obtained from the spectral broadening of the bright exciton resonance on the order of 25\,meV from exciton-phonon interaction\,\cite{Selig2016,Raja2018,Brem2019}.
As discussed in the literature also for TMDCs\,\cite{Mouri2014,Kulig2018}, one could approximately use the relation $D=k_BT(\tau_s/m_X)$ to estimate the diffusion coefficient, where $k_B$ is the Boltzmann constant, $T$ is the exciton temperature, and $m_X$ is the total mass of the exciton.
Due to the fast cooling of excitons expected at elevated temperatures\,\cite{Selig2018} compared to much longer population lifetimes, the exciton temperature should be close to that of the lattice.
Depending on the choice of the mass parameter $m_X$, e.g., between 0.5 and 1 of the free electron mass $m_0$, the scattering time of 30\,fs yields diffusion coefficients between 1 and 2 cm$^2$/s.
Notably, this simple relation severely \emph{underestimates} measured values of the diffusion coefficient in the range of 5 to 10\,cm$^2$/s [c.f. Fig.\,\ref{fig2}\,(b)], motivating a more detailed analysis.

\begin{figure}[t]
	\centering
			\includegraphics[width=8.0 cm]{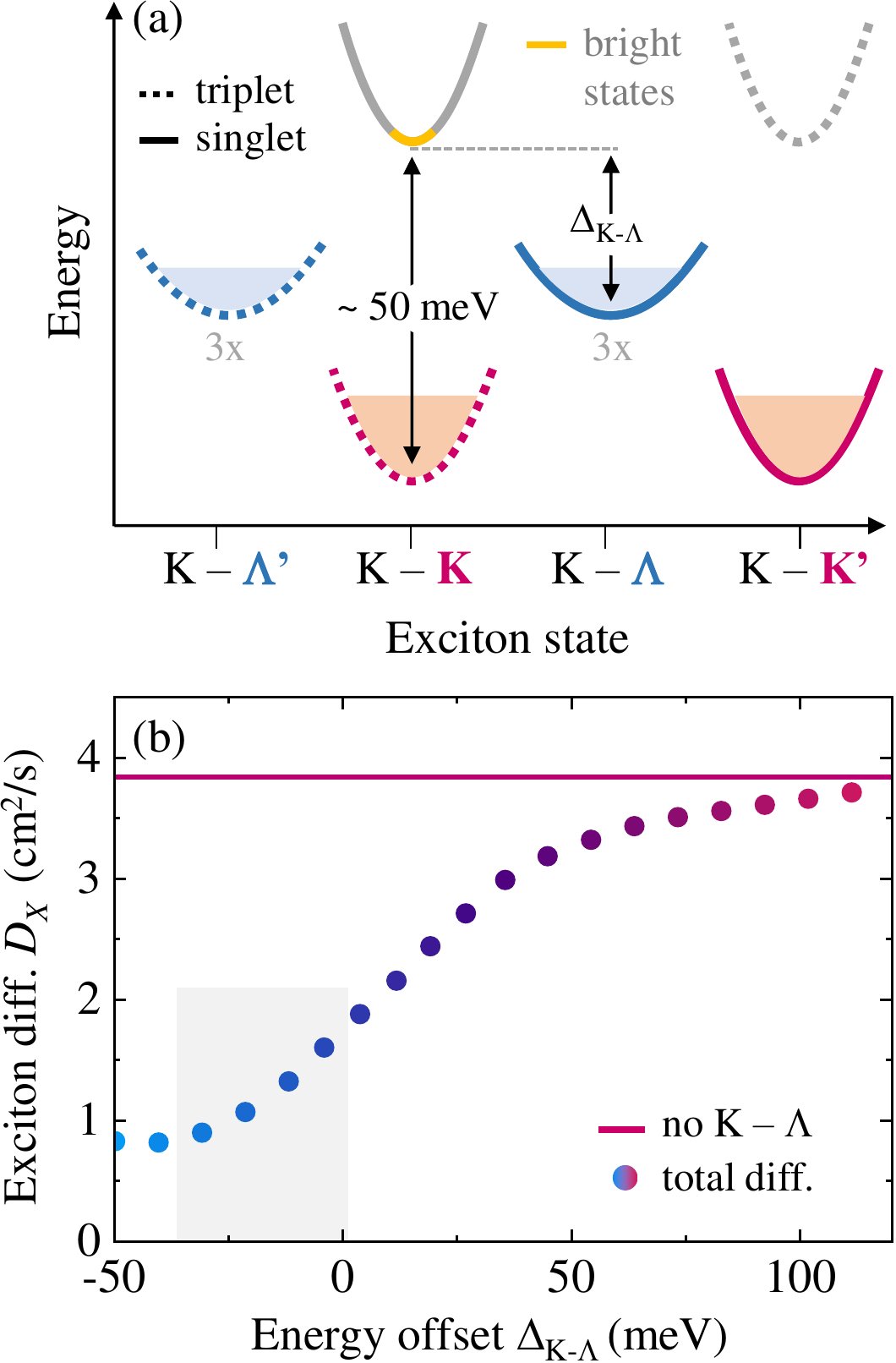}
		\caption{(a) Schematic illustration of the exciton bandstructure in hBN-encapsulated WS$_2$ relevant for the studied diffusion at room temperature.
		The exciton states are denoted according to their respective electronic transitions, including only the ones originating from the top valence band at K-valley due to a large energy-offset of the lower spin-split band.
		The expected occupancy for the majority of the exciton population at room temperature is approximately indicated by the filled areas.
		The energy offset of the threefold degenerate $K-\Lambda$ and $K-\Lambda'$ states is denoted by $\Delta_{K-\Lambda}$.		
		(b) Theoretically calculated total exciton diffusion coefficients at room temperature as function of the energy offset of the $K-\Lambda$ exciton relative to the bright $K-K$ singlet state.
		The exciton diffusion calculated including only the $K-K$ and $K-K'$ states is added for comparison.
		Gray area indicates the estimated splitting of the $K-\Lambda$ exciton states based on the conduction band offsets from Ref.\,\onlinecite{Kormanyos2015}.
		}
	\label{fig3}
\end{figure}

The initial step is to accurately consider the role of the dark exciton states that are particularly important in tungsten-based TMDCs such as the studied WS$_2$ monolayers.
Both the spin-splitting of the conduction band and the presence of the $\Lambda$ valley (also labeled as $Q$ in the literature) between $\Gamma$ and $K$ combined with influence of the Coulomb interaction results in a manifold of exciton states that are relevant at room temperature.
These can be labeled by the respective electron transitions: \textit{intra}valley $K-K$ singlets and triplets, \textit{inter}valley $K-K'$ and $K-\Lambda$ singlets, and $K-\Lambda'$ triplets as well as the corresponding permutations for the electronic transitions from the $K'$ valley. 
The terms ``singlet'' and ``triplet'' relate to the conventional electron-hole picture and correspond to the total exciton spin of 0 (electron and hole spins are antiparallel) and 1 (electron and hole spins are parallel), respectively.
The relevant parts of the exciton bandstructure for monolayer WS$_2$ are schematically illustrated in Fig.\,\ref{fig3}\,(a).

In the experiment, the detected luminescence arises from the bright exciton states at $K-K$ and $K'-K'$ with the center-of-mass momentum close to zero, i.e., those within the radiative cone.
The majority of the population, however, that also determines the diffusion dynamics on the studied time-scale of several 10's of picoseconds resides in lower-lying states\,\cite{Rosati2019, Brem2020}.
The energy separation of intravalley $K-K$ triplets and intervalley $K-K'$ singlets with respect to the bright $K-K$ state is well established for W-based TMDC monolayers\,\cite{Zhang2017,Wang2017}.
Taking the single-particle conduction-band splitting\,\cite{Kormanyos2015} and electron-hole interaction into account by solving the Wannier equation in the thin-film limit, one obtains the corresponding splitting of about 50\,meV for hBN-encapsulated WS$_2$.
Here we also note that as long as the value is larger than the thermal energy at room temperature of about 26\,meV, these exciton states should be predominantly populated.
In contrast, the precise energy offset for the $K-\Lambda$ states is less clear due to more challenging experimental access. 
Recently, conduction band offsets for MoSe$_2$ and WSe$_2$ were reported that were obtained for electron doping in the 10$^{13}$\,cm$^{-2}$ range\,\cite{Zhang2019}.
The values that we use for WS$_2$, however, rely on the conduction band offsets obtained from ab-initio calculations that can vary within many 10's of meV for WS$_2$ in particular\,\cite{Kormanyos2015, Deilmann2019, Malic2018}.

In the context of the band structure depicted in Fig.\,\ref{fig3}\,(a), the  symmetry between triplet and singlet states with the same energy and effective masses results in the same scattering and diffusion efficiency.
As a consequence, the overall exciton population in triplet states diffuses in the same way as the population in the corresponding singlet states. 
In addition, the impact of spin-flip exciton-phonon scattering on the overall diffusion should be very small due to the long scattering times\,\cite{Song2013} compared to the ultrafast spin-conserving scattering\,\cite{Selig2016}. 
We can thus neglect spin-flip processes in our model. 
In this situation, the phonon modes involved in the exciton scattering are the in-plane longitudinal and transversal acoustic (LA, TA) and optical (LO, TO) modes and the symmetric out-of-plane A$_1$ mode at the $\Gamma$, $\Lambda$, $M$, and $K$ points\,\cite{Jin2014}.

To microscopically calculate exciton diffusion we scan across a relevant range of the $K-\Lambda$ exciton energy splitting with respect to the bright $K-K$ exciton state and compute the total diffusion of the distributed exciton population.
Following our previous work\,\cite{Rosati2019}, we set up equations of motion for the exciton occupation in different valleys by exploiting the Heisenberg equation with the many-particle Hamilton operator in the excitonic basis, including carrier--light, carrier--phonon, and carrier--carrier interactions. 
We use ab-initio parameters for the electronic\,\cite{Kormanyos2015} and phononic\,\cite{Jin2014} properties of monolayer WS$_2$ and calculate the time-, space- and energy-resolved dynamics of the exciton occupation in the Wigner representation. 
We then extract the exciton diffusion coefficient from the slope of the temporal evolution of the squared width of the exciton spatial distribution.

Theoretically computed exciton diffusion coefficients are presented in Fig.\,\ref{fig3}\,(b) as function of the energy offset of $K-\Lambda$ \textit{exciton} states with respect to the bright $K-K$ states.
Intra- and intervalley scattering processes are included for the $K-K$, $K-K'$, and $K-\Lambda$ valley configurations and the diffusion coefficient of the overall exciton population is extracted from the numerical simulations.
As a limiting case, we also include the result for the exciton diffusion coefficient while neglecting $K-\Lambda$ states for comparison.
In addition, the approximate range of energy positions of the $K-\Lambda$ exciton state based on different ab-initio values of the conduction band splitting of the $\Lambda$ valley from single-particle bandstructures \cite{Kormanyos2015} are indicated by the gray area.
Altogether, our theoretical results show that one can, in principle, expect exciton diffusion coefficients in the range of 1 to 4 cm$^2$/s, depending on the relative energy of the $K-\Lambda$ state.
As a consequence of a complex exciton bandstructure, the diffusion coefficients can thus indeed deviate from the initial simplified estimations that were based only on the scattering time of the bright state.
Nevertheless, the values are still below experimentally obtained coefficients in the 5 to 10\,cm$^2$/s range, especially for more realistic splitting parameters of the $K-\Lambda$ exciton.

Here we emphasize again that we can largely exclude potential contributions from non-thermalized hot excitons, since the room temperature cooling should occur extremely fast, within the first few picoseconds\,\cite{Selig2018, Rosati2019} in contrast to the studied range of many 10's of picoseconds.
In addition, measurements of consistently high, constant diffusion across roughly an order of magnitude of excitation densities indicate that hot-phonon related phenomena\,\cite{Leo1988,Glazov2019,Perea-Causin2019} should not play a role under these conditions.
Since the finding of diffusion coefficients (or effective mobilities) that are much \textit{higher} than the theoretical limit is rather uncommon, we need to consider alternative sources that may facilitate efficient propagation of the optical excitations.
In the following we argue that a finite population of free electron-hole plasma with comparatively high diffusion coefficients can indeed provide such a source.

\subsection {Influence of the electron-hole plasma}
\label{linear-B}

In TMDC monolayers it is commonly understood that the excitons are stable under a broad range of density and temperature conditions due to their high binding energies\,\cite{Wang2018}.
While it is indeed often the case, there is nevertheless a number of scenarios where large populations of free electrons and holes can be expected\,\cite{Steinhoff2017}.
For example, at high excitation densities beyond the Mott-transition threshold, the excitons cease to be bound states due to screening and Pauli-blocking\,\cite{Shah1977,Snoke2008,Steinhoff2017,Chernikov2015c} leading to the emergence of a dense electron-hole plasma.
Interestingly, also at sufficiently \emph{low} densities the occupation of free carrier states can be favored and the formation of the excitons suppressed.

This phenomenon, known as ``entropy ionization'' from the literature\,\cite{Mock1978}, can be qualitatively understood by considering the free energy $\mathcal F=\mathcal U-\mathcal S T$ of the system that is minimized at thermal equilibrium.
The exciton contribution to the internal energy $\mathcal U$ is lower than that of the free carriers due to the finite binding energy, favoring their formation at lower temperatures $T$.
On the other hand, the entropy $\mathcal S$ is higher for an unbound electron-hole pair in contrast to the exciton due a larger number of micro-states available for uncorrelated charge carriers.
At elevated temperatures and sufficiently \textit{low} densities the entropy term becomes dominant, leading to the preferential occupation of free electron and hole states instead of the excitons.

For TMDC monolayers at room temperature, it was predicted\,\cite{Steinhoff2017} that one can expect significant population of free carriers already for electron-hole pair densities of 10$^{10}$\,cm$^{-2}$ and below, depending on the dielectric environment determining the value of the exciton binding energy.  
These densities can be relevant for a broad range of realistic experimental condition both for continuous wave and pulsed excitation, as used in our study, motivating the necessity to examine the exciton-plasma equilibrium more closely.
In the following, we use a simplified approach of neglecting any Coulomb renormalization effects and treating the excitons and free charge carriers in our system as classical gases.
This should be well justified by comparatively high temperatures and low occupation densities between 10$^{8}$ and 10$^{10}$\,cm$^{-2}$ in the studied excitation regime ($k_B T \gg n_p/\mathcal D_p$, see below).
We estimate that the upper applicability limit should be on the order of several 10$^{12}$\,cm$^{-2}$ at room temperature.

The free energy density can be then presented in the following form:
\begin{equation}
\label{freeE}
\mathcal F=\sum_p{n_p k_BT\,\left(\ln\frac{n_p}{\mathcal D_p k_BT}-1\right)+n_pE_p},
\end{equation}
where $n_p$ is the density of the $p$-th type of particle with the subscript $p$ running through all relevant free electron and hole states at $K$, $K'$ and $\Lambda$ valleys as well as the excitons, taking into account the respective degeneracies.
$E_p$ denotes the relative energy of the state including band offsets and binding energies, and $\mathcal D_p$ represents the density of states in two-dimensions, according to:
\begin{equation}
\mathcal D_p=g_p\frac{m_p}{2\pi\hbar^2},
\end{equation}
with the center-of-mass $m_p$ and the degeneracy factor $g_p$.
In addition, we set conditions for the conservation of the total, photoinjected electron-hole pair density $N_{tot}$ in the system:\footnote{For the estimation of the exciton-plasma equilibrium of photoexcited charge carriers we disregard any thermal activation of electrons and holes across the band gap.}
\begin{equation}
N_{tot}=\sum_{\stackrel{k}{(excitons)}}{n_{X,k}}+\sum_{\stackrel{j}{(holes)}}{n_{h,j}}\equiv\bar{n}_X+\bar{n}_h,
\end{equation} 
and for the charge neutrality:
\begin{equation}
\label{neutr}
0=\sum_{\stackrel{j}{(holes)}}{n_{h,j}}\,-\sum_{\stackrel{i}{(electrons)}}{n_{e,i}}\equiv\bar{n}_h-\bar{n}_e.
\end{equation} 
Here, the densities of the individual exciton ($n_X$), hole ($n_h$), and electron ($n_e$) states are enumerated by $k$, $j$, and $i$, respectively.
The bar on top denotes the total density. 
These expressions are derived under the conditions of negligible doping in the system (see the discussion below and in the Appendix\,\hyperlink{AppA}{A}). 
From the Eqs.\,\eqref{freeE}\ldots\eqref{neutr} we obtain a set of equations for the individual densities $n_{p}$ that are solved numerically as function of the total density $N_{tot}$ for a fixed temperature.
The equations and their solutions are presented in detail in Appendix\,\hyperlink{AppA}{A}.

Interestingly, one can also show that the results for any complex multi-band scenario with parabolic bands can be obtained from a single equation for the \textit{total} densities of excitons $\bar{n}_X$, free holes $\bar{n}_h$, and free electrons $\bar{n}_e$ (commonly known as Saha-equation or mass action law\,\cite{Mock1978}) with an appropriately chosen parameter $S$:
\begin{equation}
\label{Saha}
\frac{\bar{n}_e\bar{n}_h}{\bar{n}_X}=S,
\end{equation}
and the analytical solution for the exciton fraction $\alpha_X$:
\begin{equation}
\label{SahaSol}
\alpha_X \equiv \frac{\bar{n}_X}{N_{tot}}=1+\frac{S}{2N_{tot}}-\sqrt{\left(\frac{S}{2N_{tot}}\right)^2+\frac{S}{N_{tot}}}.
\end{equation}
The derivation of the mass action law from Eq.\,\eqref{freeE} and the comparison between the full multi-band solution and the Saha-equation are given in Appendix\,\hyperlink{AppA}{A}. 

\begin{figure}[t]
	\centering
			\includegraphics[width=8.0 cm]{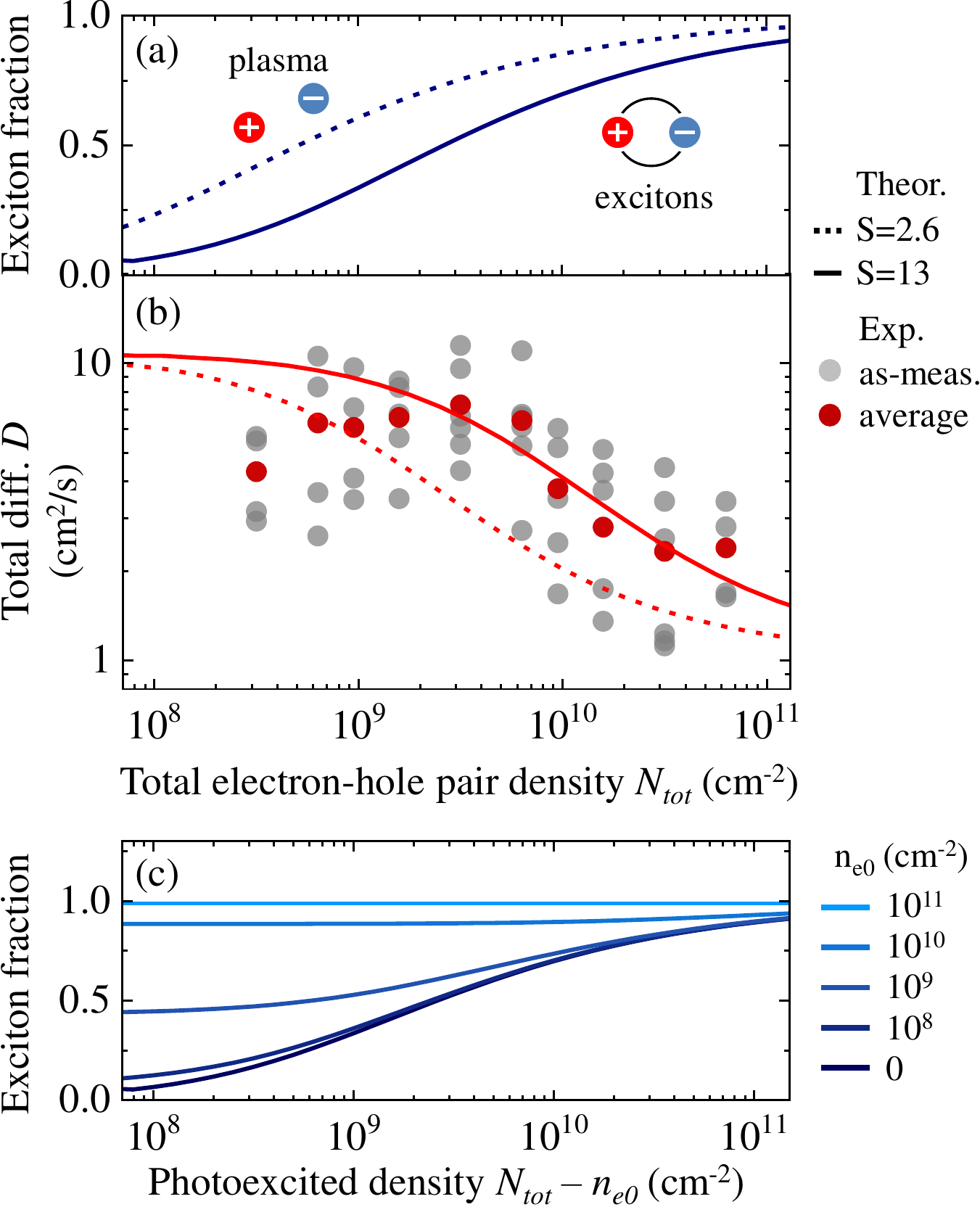}
		\caption{(a) Calculated fraction of the combined exciton population as function of the total electron-hole pair density for two different values of the Saha-parameter $S$, given in units of 10$^8$\,cm$^{-2}$.
		(b) Corresponding density dependence of the composite diffusion coefficient of propagating excitons and free charge carriers.
		Experimental data in the low-excitation regime from Fig.\,\ref{fig2}\,(b) are presented for direct comparison.
		(c) Calculated exciton fractions for varying electron doping densities as function of the photoexcited electron-hole pair density.
		}
	\label{fig4}
\end{figure}

Using the bandstructure parameters from the literature and evaluating the exciton energies of the studied hBN-encapsulated WS$_2$ monolayers, we obtain $S$=2.5\,$\times$\,10$^8$\,cm$^{-2}$.
The corresponding exciton fraction is presented in Fig.\,\ref{fig4}\,(a) by the dotted line.
To provide an example illustrating the sensitivity with respect to the key parameters we also show the result for slight adjustments of the fixed values (within 10\% to 30\%): $\Delta m_X$\,=\,+0.3\,$m_0$, $\Delta E_{bind}$\,=\,$-20$\,meV, $\Delta T$\,=\,+15\,K, $\Delta E_{\Lambda}$\,=\,+23\,meV.
This yields $S$=13\,$\times$\,10$^8$\,cm$^{-2}$ and the corresponding exciton fraction is shown by the solid line in Fig.~\ref{fig4}\,(a).
In both cases the exciton fraction strongly decreases below the total density of 10$^{10}$\,cm$^{-2}$, with the majority of the population forming free electron-hole plasma.
Hence, it is indeed necessary to examine the diffusion properties of a \textit{composite} system in the presence of sizable densities for both bound and unbound charge carriers.

For this purpose, we consider a set of two coupled diffusion equations for the total densities of excitons $\bar{n}_X$ and electron-hole plasma $\bar{n}_{eh}$, since free electrons and holes should diffuse together to conserve electro-neutrality:
\begin{subequations}
\label{coupled}
\begin{align}
\frac{\partial \bar{n}_X}{\partial t} = D_X \Delta_{\bf{r}} \bar{n}_X+D_X' \Delta_{\bf{r}} \bar{n}_{eh} \notag \\ - \frac{\bar{n}_X}{\tau_X} - \frac{\bar{n}_X - \nu_X \bar{n}_{eh}}{\tau_i},\\
\frac{\partial \bar{n}_{eh}}{\partial t} = D_{eh} \Delta_{\bf{r}} \bar{n}_{eh} + D_{eh}' \Delta_{\bf{r}} \bar{n}_{X}\notag \\ - \frac{\bar{n}_{eh}}{\tau_{eh}} + \frac{\bar{n}_X - \nu_X \bar{n}_{eh}}{\tau_i}.
\end{align}
\end{subequations}
Here $D_X$ is the exciton diffusion coefficient and $D_{eh}$ is the electron-hole plasma diffusion coefficient, i.e., the average diffusion coefficient of the free electrons and holes.
$D_X'$ and $D_{eh}'$ are the trans-diffusion coefficients related to the exciton current driven by the plasma gradient and vice-versa.
The recombination times of excitons and electron-hole pairs in the plasma are denoted by $\tau_x$ and $\tau_{eh}$, respectively.
The exciton ionization rate is represented by $\tau_i$ and $\nu_X \equiv \bar{n}_X^{(eq)}/\bar{n}_{eh}^{(eq)}$ is the ratio of the exciton density to the density of unbound pairs in thermal equilibrium.
The Laplace operator is denoted by $\Delta_{\bf{r}}$, defined with respect to the radial in-plane coordinate $\bf{r}$.

In the following we assume that the ionization equilibrium is established sufficiently fast as compared to the population lifetime, i.e., $\tau_i \ll \tau_X,\tau_{eh}$. 
On the other hand, $\tau_i$ should be slower than the total momentum relaxation times of excitons and of plasma, since the later include both processes associated with $\tau_i$ but also additional exciton-phonon and electron/hole-phonon scattering events that are very efficient in TMDCs at room temperature.  
For simplicity we further assume that the densities of excitons and of electron-hole pairs are sufficiently low so that plasma-driven exciton currents and exciton-driven plasma currents can be neglected.
Hence we set $D_X'$=$D_{eh}'$=0 in the following and consider only the diffusion of the individual constituents $D_X$ and $D_{eh}$. 
\footnote{Trans-diffusion is related to the collisions between the free charge carriers and the excitons. 
The rate of these collisions is negligible compared to the rate of phonon-assisted processes at low densities studied here.}

As a consequence, in the equation set~\eqref{coupled} there is a fast mode associated with the relaxation towards the equilibrium $\bar{n}_X - \nu_{X} \bar{n}_{eh}$ that quickly relaxes to zero and a slow mode corresponding to the total density of the particles $N_{tot} = \bar{n}_X + \bar{n}_{eh}$ that is relevant for the combined diffusion.
Following from the relations $\bar{n}_{eh} = N_{tot}/(1+\nu_x)$ and $\bar{n}_{x} = \nu_x N_{tot}/(1+\nu_x)$ for fast ionization processes, the total particle density $N_{tot}$ obeys a simple diffusion equation:
\begin{equation}
\label{diffusion:n}
\frac{\partial N_{tot}}{\partial t} = \Delta(\bar D  N_{tot}) - \frac{N_{tot}}{\bar \tau}
\end{equation}
with
\begin{subequations}
\begin{align}
\label{D:eff}
\bar D = \alpha_{eh} D_{eh} +  \alpha_X D_X, \\
\bar \tau^{-1} = \alpha_{eh} \tau_{eh}^{-1}  +  \alpha_X \tau_{X}^{-1} .
\label{time:eff}
\end{align}
\end{subequations}
Here, $\alpha_{eh}$=$\bar{n}_{eh}/(\bar{n}_X+\bar{n}_{eh})$ and $\alpha_X$=$\bar{n}_X/(\bar{n}_X+\bar{n}_{eh})$ denote the respective fractions of plasma and excitons.
It follows that the combined diffusion coefficient of the electron-hole population is composed of the individual coefficients of plasma and excitons weighted by their relative densities. 
Here we emphasize that both fractions $\alpha_{eh}=\alpha_{eh}[N_{tot}(\bm r, t),T]$ and $\alpha_{X}=\alpha_X[N_{tot}(\bm r, t),T]$ are functions of temperature and local density, rendering the diffusion equation~\eqref{diffusion:n} generally non-linear.

To describe the experimentally studied scenario we then combine the description of exciton-plasma equilibrium with that of the multi-component diffusion. 
The individual diffusion coefficients of excitons and plasma are fixed to the results of the microscopic calculation via Bloch-equations approach outlined further above and to the ab-initio calculations from Ref.\,\onlinecite{Jin2014} respectively, i.e. to $D_X$=1\,cm$^2$/s and $D_{eh}$=11\,cm$^2$/s (using literature parameters for the electron and hole properties as well as $\Delta_{K-\Lambda} = - 26$\,meV).
As an initial condition at time $t=0$ we fix the total density distribution to the excitation spot profile and the estimated injected densities from the experiment.
Subsequently, we solve Eq.~\eqref{diffusion:n} numerically with $\bar D$ depending on $N_{tot}$ via Eqs.~\eqref{D:eff} and \eqref{SahaSol} and the population lifetime calculated after Eq.~\eqref{time:eff} using a similar approach. In this way we obtain $N_{tot}(\bm r,t)$.
From these we extract an effective diffusion coefficient for each excitation density to compare with the experiment.  

Theoretically obtained density-dependence of the effective diffusion coefficient is presented in Fig.\,\ref{fig4}\,(b).
The results correspond to the two choices of the Saha-parameter shown in Fig.\,\ref{fig4}\,(a) and include the experimental data both from the individual measurements and the average.
Considering the spread of the data and potential variations of the system parameters, the quantitative agreement is very reasonable.
Importantly, the model captures both the high values of the diffusion coefficient found at low densities but also naturally explains the subsequent decrease observed at higher densities, marked by the distinct regimes (1) and (2) in Fig.\,\ref{fig2}\,(b), respectively.
We may thus conclude that sizable populations of the electron-hole plasma in the low-density regime due to entropy ionization of the excitons can indeed drive the overall propagation efficiency of the photoexcited charge carriers.
As the density increases, the relative fraction of plasma decreases and the diffusion coefficient approaches the intrinsic value for the excitons.
We also emphasize that the reduced value of the exciton binding energy in hBN-encapsulated samples is the main reason to expect entropy ionization in the studied density range. 
In as-exfoliated samples on SiO$_2$ substrates, however, the exciton binding energies are more than 100\,meV higher than in the encapsulated ones due to reduced dielectric screening so that entropy ionization should occur at much lower densities (see Fig.\,\ref{figA0} in Appendix \hyperlink{AppA}{A}).

We note, however, that the proposed model is subject to an important limitation. 
Realistic TMDC materials, even after encapsulation in hBN, often exhibit a small residual amount of doping. 
It is usually relatively low, as in the case of the studied samples, where we can only estimate it to be of $n$-type and below 10$^{11}$\,cm$^{-2}$ from reflectance measurements at cryogenic temperatures.
To analyze the implications of doping, the above analysis can be readily extended to allow for the presence of resident carriers (see Appendix\,\hyperlink{AppA}{A} for details).
Calculated exciton fractions for the doping densities between 10$^{8}$ and 10$^{11}$\,cm$^{-2}$ are presented in Fig.\,\ref{fig4}\,(c) as function of the photoexcited electron-hole-pair density. 
We see that already for relatively low densities on the order of 10$^{9}$\,cm$^{-2}$, the equilibrium is shifted to favor the exciton population. 
A finite doping density of free electrons would thus stabilize the exciton formation.
One can roughly understand that by considering that it is much easier for a photoexcited hole to find an electron to form an exciton in the presence of resident electrons.

Importantly, such free-carrier densities can not be unambiguously excluded in the experiment, even if one might expect non-uniform distribution of dopants, localized in smaller puddles and leaving undoped areas in-between governing the diffusion.
In addition to that, bright and dark trion states should form and further affect the specific ratios of the electron-hole pair distributions in the equilibrium.
Here, we note that the additional formation of trion states is omitted in the simplified analysis presented above and in Fig.\,\ref{fig4}\,(c) that is aimed to illustrate the general influence of residual carriers on the equilibrium  between free and bound electron-hole pairs.
However, the correction to the total internal energy of a trion state of about 20 to 30\,meV is rather small (about 10$\%$) in comparison to that of the neutral exciton on the order of 200\,meV.
Thus, we do not expect the free carrier fraction to be strongly affected by redistribution of the exciton population between the neutral and charged composite states.
Moreover, significant populations of the trions could affect the diffusion in the linear and non-linear regimes.
The higher total mass of the trion should already slow down the propagation, similar to the observed interplay between trions and free carriers\,\cite{Lui2014} as well as affect the scattering rates, in particular with the phonons.
While the emission in the studied samples is dominated by neutral excitons, the simplified analysis presented in Fig.\,\ref{fig4}\,(c) should be still considered as an illustration of the general influence of residual doping on the equilibrium exciton fractions.
The actual experimental scenario is likely to be more complex, strongly motivating further investigations into the subject.
These would ideally combine precise determination and control of the doping down to very low densities as well as direct observations of the electron-hole plasma by intraband, interband, or photocurrent spectroscopy. 

To summarize, the following can be concluded with respect to the initial question regarding the mechanisms determining room-temperature linear diffusion in the studied hBN-encapsulated WS$_2$ monolayers.
In general, it is necessary to consider the presence of dark states in tungsten-based materials for accurate description of the exciton propagation.
Importantly, the resulting effective exciton diffusion coefficient can strongly vary depending on the relative valley alignment in the conduction band.
The exciton diffusion alone, however, does not seem to fully account for the experimental findings, \textit{underestimating} the measured values by a factor of at least 2 and up to 10.
A promising hypothesis involves the presence of sizable plasma populations in the system due to the entropy ionization of excitons at low excitation densities, providing a reasonable quantitative description of the experiment.
Photoexcited, thermally stabilized free electrons and holes are expected to diffuse much faster than the excitons.
They can thus dominate the propagation of the total photoexcited population, even as the role of the residual doping remains an important consideration for the validity of this scenario.
 
\subsection {Recombination dynamics}
\label{linear-C}

\begin{figure}[t]
	\centering
			\includegraphics[width=8.4 cm]{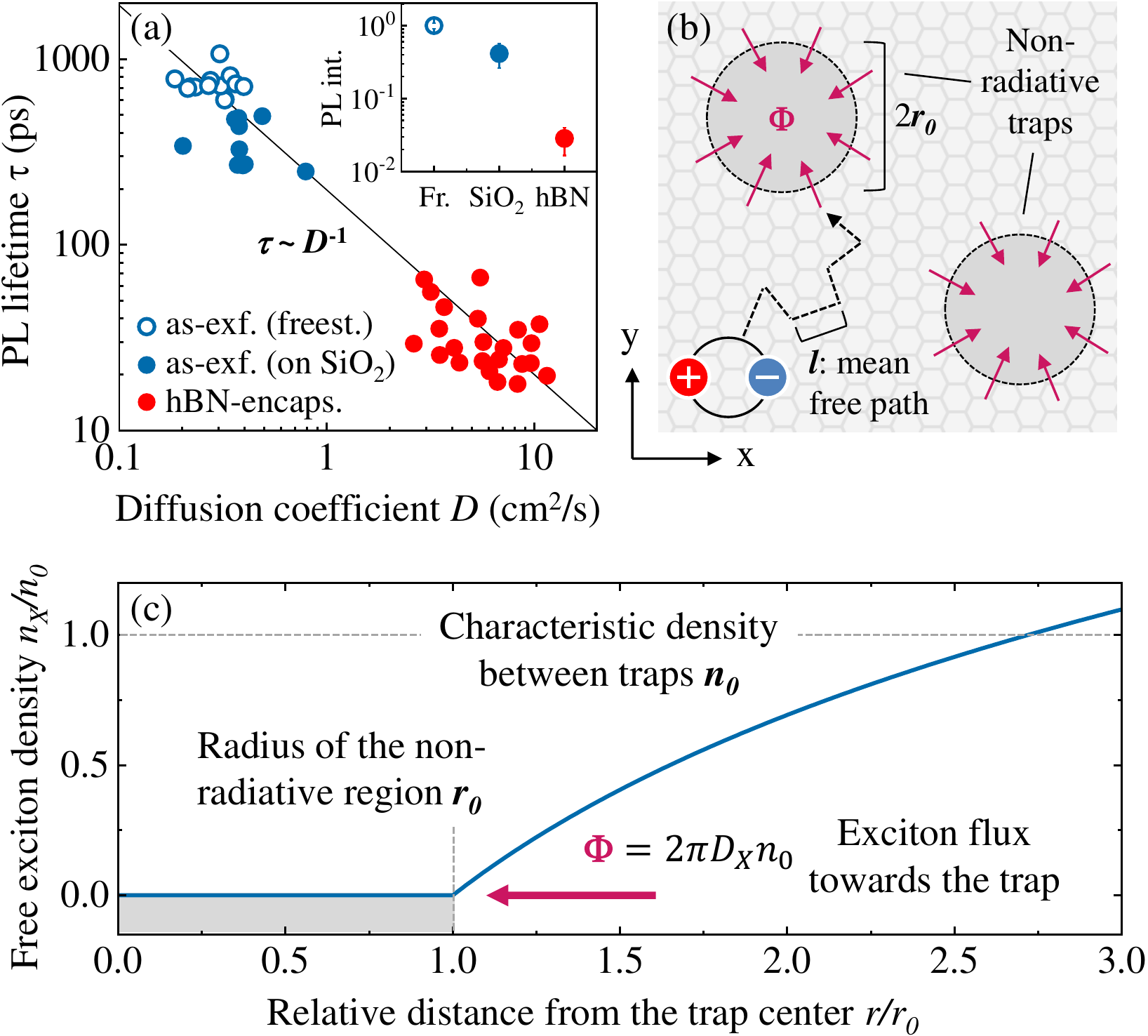}
		\caption{(a) Measured PL lifetimes of as-exfoliated and hBN-encapsulated WS$_2$ monolayer samples as function of the experimentally obtained diffusion coefficients in the linear regime.
		The as-exfoliated samples include those placed on SiO$_2$/Si substrates and freestanding flakes were suspended across 5$\times$5\,$\mu$m$^2$ holes cut into SiN membranes (see Ref.\,\onlinecite{Kulig2018} for details).
		The inset shows the comparison of total PL intensities measured for constant excitation power in the linear regime.
		(b) Schematic illustration of a propagating exciton in a drift-diffusion model reaching a region facilitating non-radiative recombination.
		(c) Free exciton density in the vicinity of non-radiative trap region obtained from the model according to the Eq.\,\eqref{axial}.
		}
	\label{fig5}
\end{figure}

In the previous subsection we discussed possible origins of the observed efficient diffusion in hBN-encapsulated samples.
Notably, in addition to determining the diffusion coefficients we also find a strong correlation between rapid propagation and carrier lifetime dynamics that we address in the following.
As we demonstrate, the experimental findings are rather robust across many sample positions and individual flakes both in our data and in the literature\,\cite{Hoshi2017}.
The relationship between population lifetime and diffusion coefficients is presented in Fig.\,\ref{fig5}\,(a).
Here, the PL lifetime is plotted as function of the linear diffusion coefficient for a sufficiently large number of studied flakes and measurement positions.
The data illustrates consistent differences between the as-exfoliated (on SiO$_2$/Si substrates and freestanding) and hBN-encapsulated samples as well as within the individual sample sets.

Considering the underlying mechanisms, room-temperature recombination is typically limited by non-radiative processes associated with defects.
This is further supported by measurements of the total PL intensity which decreases for the sample sets with shorter lifetimes on a similar scale as the lifetimes themselves, as illustrated in the inset of Fig.\,\ref{fig5}\,(a).
Thus, a number of possible scenarios involving changes in the exciton trapping efficiency and a variety of non-radiative recombination pathways may occur as a consequence of encapsulation.
It is nevertheless instructive to examine the following basic model.
As a limiting case, we can assume that the non-radiative recombination is provided by a finite trap density that is of the same order of magnitude in as-exfoliated and encapsulated samples.
This proposition could be reasonable considering that many of the trap states are likely to already reside within the material itself\,\cite{Schuler2019}, at least in the studied WS$_2$ samples, and are not necessarily associated with the choice of the substrate.
Only moderate increase of the exciton lifetime from about 0.5...0.7\,ns in SiO$_2$-supported samples to 0.8...1\,ns in freestanding flakes\,\cite{Kulig2018} are consistent with this assumption.

Since the excitons have to propagate and reach the regions of non-radiative traps to be captured, as schematically illustrated in Fig.\,\ref{fig5}\,(b), we could expect that the capture rate may be indeed dependent on the diffusion efficiency.
That would mean that the faster the excitons or free charge carriers travel across the material, the sooner they will encounter non-radiative centers and recombine.
For a quantitative description of the diffusion-limited non-radiative capture we extend the model of the carrier capture used for bulk materials~\cite{abakumov_perel_yassievich} to two-dimensional semiconductors. 
The details of the model are presented in the Appendix~\ref{AppB} and we summarize the main results in the following.

Here we consider the case where the characteristic radius $r_0$ of the traps is large compared with the mean free path $\ell$ of the excitons, i.e., $r_0 \gg \ell = \langle v \rangle\tau$ with $\tau$ being exciton momentum relaxation time and $\langle v\rangle$ being the exciton thermal velocity. 
We note that this approach works analogously both for excitons and plasma diffusion. 
In order to determine the capture rate we solve the diffusion equation for the steady state (setting $\partial n_{X}/\partial t=0$) in the vicinity of the trap:
\begin{equation}
\label{diffusion:trap}
D \Delta n_X=0, 
\end{equation}
 with the boundary condition 
 \begin{equation}
 \label{boundary}
 n_X(r_0)=0. 
 \end{equation}
The condition~\eqref{boundary} means that all excitons which reach the trap are captured and recombine non-radiatively.
In the axially symmetric geometry the solution of Eqs.~\eqref{diffusion:trap} and \eqref{boundary} takes the form
\begin{equation}
\label{axial}
n_X(r) = n_0 \ln{(r/r_0)}, \quad r>r_0,
\end{equation}
where $n_0$ is the characteristic density of excitons between the traps.
The radial dependence of the free exciton density $n_X(r)$ from Eq.\,\eqref{axial} is illustrated in Fig.\,\ref{fig5}\,(c).
We note that formally the density $n(r)$ would diverge at $r\to \infty$.
This is, however, an artifact of the two-dimensional case. 
More importantly, for realistic scenarios there are always other traps at finite distances that would compensate the increase of $n_X (r)$. 

The total flux $\Phi$ of the excitons towards the trap is then readily found from the continuity equation
\begin{equation}
\label{flux}
|\Phi| =  2\pi r_0 D_X \left.\frac{d n}{dr}\right|_{r=r_0} = 2\pi D_X n_0,
\end{equation}
obtaining the recombination rate from the flux according to:
\begin{equation}
\label{tau-1:1}
\frac{1}{\tau_r} \equiv \frac{N_{tr} \Phi}{n_0} = 2\pi D_X N_{tr}.
\end{equation}
Here $N_{tr}$ represents the density of non-radiative recombination traps, so that $1/(D_XN_{tr})$ is associated with the typical time which takes for exciton to diffuse between the traps.

Notably, Eq.\,\eqref{tau-1:1} corresponds to the experimentally observed correlation $\tau_r\propto D_x^{-1}$, see Fig.~\ref{fig5}(a), and could thus be a reasonable interpretation of our findings.
For the typical parameters of the studied hBN-encapsulated WS$_2$ monolayer samples in the low density regime, $\tau_r$=25\,ps and $D$=8\,cm$^2$/s, the \textit{effective} density of the non-radiative trap regions would be on the order of 10$^{9}$\,cm$^{-2}$.
The corresponding average separation between the traps of several 100's of nanometers would then limit the diffusion length in the studied samples.
Here we emphasize that, according to our model, $N_{tr}$ represents an \textit{equivalent} trap density assuming 100\% capture probability.
For more realistic scenarios of finite capture cross-sections, the actual density of these regions could be much higher.
Their origin could be related either to regions with high concentrations of point-defects or, alternatively, to puddles of doping\,\cite{Lien2019}, promoting non-radiative recombination.
We also note that if the effective size of an individual trap center is much smaller than the exciton mean free path, the capture time should be independent of the diffusion coefficient, as discussed in Appendix ~\ref{AppB}.
Finally, we emphasize that one can not rule out that the number of effective traps may be affected by the encapsulation, through changes either in their density or their passivation state, depending on the nature of the traps.
The analysis presented above is thus primarily intended to demonstrate that a higher diffusivity of the excitons alone can strongly impact the overall efficiency of the non-radiative capture even for a constant trap density.

\section{Non-linear regime analysis}
\label{non-linear}

In this section we discuss the regime of high excitation densities, indicated by (3) in Fig.\,\ref{fig2}\,(b), and the associated influence of Auger-like exciton-exciton scattering.
We demonstrate the effect of suppressed Auger recombination in hBN-encapsulated WS$_2$ samples on the exciton propagation and illustrate the absence of halo-like shapes in the spatial distribution of excitons as a consequence of comparatively slow Auger processes.

\subsection {Suppressed Auger recombination}
\label{non-linear-A}

\begin{figure}[t]
	\centering
			\includegraphics[width=7.5 cm]{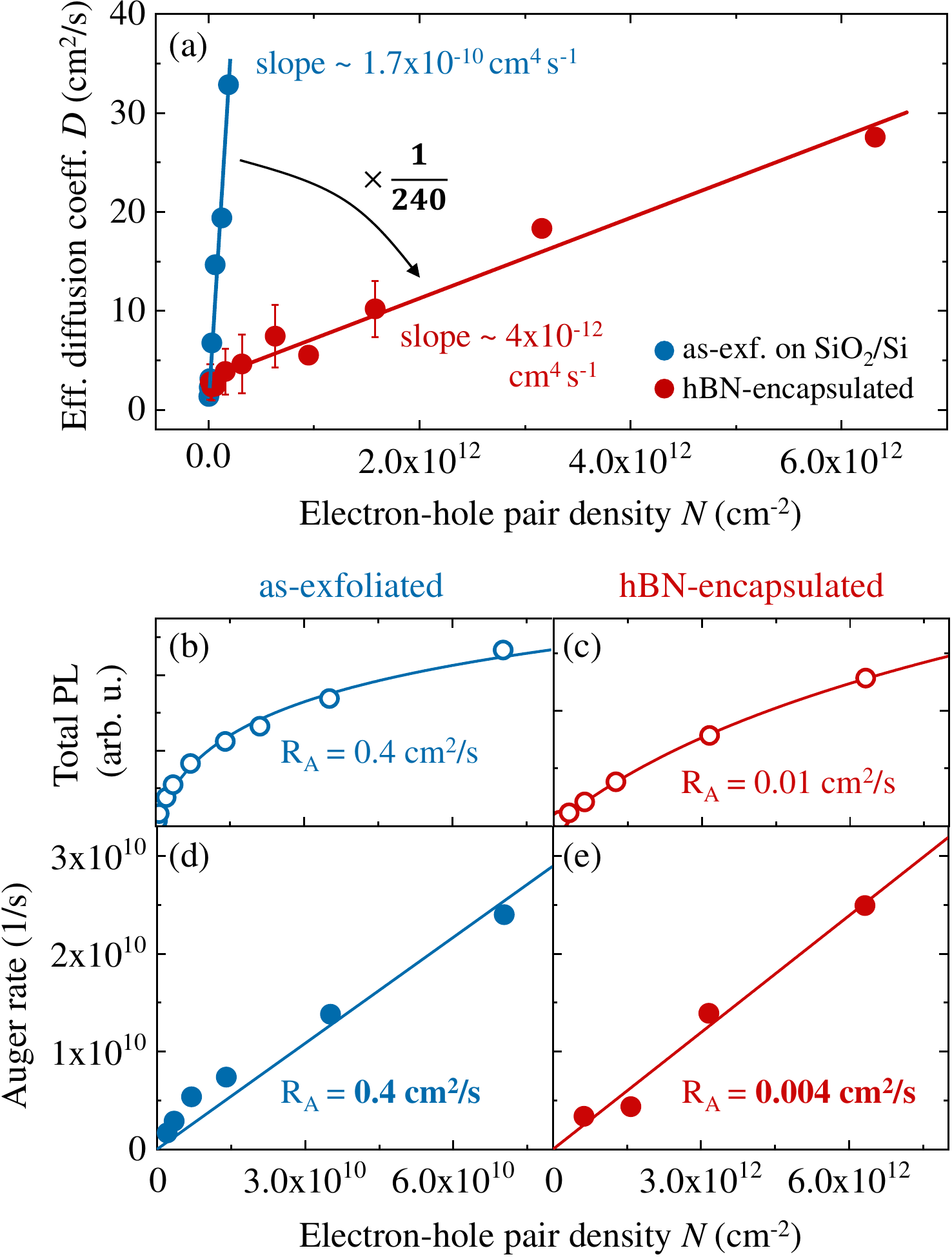}
		\caption{(a) Linear increase of the effective diffusion coefficient in hBN-encapsulated and as-exfoliated samples at elevated excitation densities (10$^{10}$ to 10$^{12}$ cm$^{-2}$). 
		The respective slopes of increasing effective $D$ and the corresponding ratio are indicated. 
		(b), (c) Auger coefficients estimated from time- and spatially-integrated total PL intensity as function of injected electron-hole pair (exciton) density for as-exfoliated and hBN-encapsulated samples, respectively. 
		The fits with a bimolecular Auger model with a fixed decay time in the low-density linear regime ($\tau_{exf}$=700\,ps and $\tau_{hBN}$=30\,ps) are shown by solid lines together with the extracted Auger coefficients $R_A$. 
		(d), (e) Extraction of the Auger coefficient from time resolved data: Density dependent increase of the recombination rate is plotted as a function of the exciton density. 
		In the bimolecular approximation, the Auger coefficient corresponds directly to the slope. 
		Note the two orders of magnitude difference in the densities on the abscissa axes for as-exfoliated (b, d) and hBN-encapsulated (c, e) data. 
		}
	\label{fig6}
\end{figure}

A key observation at high pump power conditions is a linear increase of the effective diffusion coefficient with the excitation density, presented in Fig.\,\ref{fig2}\,(b) in a double-logarithmic plot and illustrated in more detail in Fig.\,\ref{fig6}\,(a) on a linear scale. 
The absolute values of the extracted diffusion coefficients are found to increase up to 30\,cm$^2$/s and beyond both for hBN-encapsulated and as-exfoliated WS$_2$ samples.
The slope, however, is more than 200 times smaller in case of hBN-encapsulation indicating a pronounced quantitative change in the efficiency of the underlying processes with respect to the injected densities.

As demonstrated in previous studies\,\cite{Kulig2018, Perea-Causin2019}, a linearly increasing effective diffusion coefficient is attributed to the impact of non-radiative Auger-type scattering in TMDCs monolayers, commonly observed and often labeled as exciton-exciton annihilation in the literature\,\cite{Kumar2014,Mouri2014,Sun2014,Yu2016,Yuan2017,Hoshi2017,Manca2017,Binder2019,CordovillaLeon2019}. 
When two excitons interact, one of them can non-radiatively recombine and transfer the excess energy to the other that is excited to higher energy states or dissociated.
In the context of exciton diffusion, faster Auger recombination in the center of the spot leads to a flattening and additional broadening of the exciton distribution and thus to an apparent increase of the extracted diffusion coefficient\,\cite{Warren2000, Kulig2018}.
In addition, due to finite time-scales of exciton relaxation and cooling, it can cause local heating of the exciton population, contributing to increasingly rapid propagation\,\cite{Perea-Causin2019,CordovillaLeon2019}.

Interestingly, while the Auger scattering of the excitons was initially shown to be rather efficient in as-exfoliated TMDC monolayers at room temperature\,\cite{Kumar2014,Mouri2014,Sun2014}, it was found to be strongly suppressed upon hBN-encapsulation\,\cite{Hoshi2017}.
The measurements of the Auger rates in our samples fully support these observations.
The effective Auger coefficient $R_A$, defined via density-dependent recombination rate $r_A=R_A\times n_X$ in a simple bimolecular model, can be extracted either from spatially- and time-integrated total luminescence intensities as shown in Figs.\,\ref{fig6}\,(b) and (c) or from the relative increase of the PL decay rates plotted in Figs.\,\ref{fig6}\,(d) and (e)\,\cite{Kulig2018}.
The latter method should be slightly more accurate and provide more reliable numbers, since it is directly based on time-resolved traces of the emission, while the former is rather sensitive on the lifetime parameter fixed to the linear regime.
Nevertheless, for bimolecular non-radiative recombination, both approaches should lead to similar conclusions.

Altogether, we find strongly suppressed Auger recombination as a consequence of encapsulation, with $R_A$ being as low as 0.004\,cm$^2$/s in contrast to $R_A=0.4$\,cm$^2$/s for the as-exfoliated samples.
At the current stage we can only speculate about the underlying reasons of such drastic changes of $R_A$, since an accurate microscopic description of the Auger-recombination is extremely challenging.
In principle one can suggest several possible origins.
The overall reduction of the Coulomb interaction due to enhanced dielectric screening from the hBN encapsulation should already reduce Coulomb-mediated Auger scattering.
The expected decrease should roughly scale with the inverse square of the effective dielectric constant and thus be on the order of 4.
In addition, presence of disorder\,\cite{Raja2019} in as-exfoliated samples may enhance the Auger rate either due to the relaxation of the momentum conservation rule leading to a larger phase-space for scattering or due to the funneling of the excitons towards smaller local regions with lower potentials.
Alternatively, the shift in the bandstructure from the renormalization induced by additional dielectric screening from hBN would affect resonance conditions that may facilitate the efficiency of the exciton-exciton scattering, as it was discussed for WSe$_2$ monolayers\,\cite{Manca2017,Han2018}.

Most importantly, however, the reduction of the Auger coefficients by two orders of magnitude between as-exfoliated and hBN-encapsulated materials matches the observed decrease of the slope of the effective diffusion coefficients illustrated in Fig.\,\ref{fig6}\,(a).
We can thus conclude that exciton-exciton Auger recombination remains the main origin of the non-linear diffusion at elevated densities in the studied hBN-encapsulated WS$_2$ monolayers. 
Due to the strong suppression of the Auger recombination, however, the effect prominently appears at about two-orders of magnitude higher excitation densities in contrast to the as-exfoliated samples.

\subsection {Absence of exciton halos}
\label{non-linear-B}

An important additional consequence of the Auger recombination is the accumulation of excess energy in the center of the excitation spot, as briefly mentioned in the previous section.
For each non-radiative exciton-exciton scattering event, the remaining electron-hole pair gains the energy equivalent to that of the optical transition, that is about 2\,eV in case of WS$_2$.
The energy is then subsequently released in a cascade of emitted phonons promoting carrier relaxation towards the respective band minima and exciton formation. 
For sufficiently high densities, efficient Auger scattering thus leads to a local heating of the non-equilibrium phonon and exciton systems.
The associated processes are discussed in detail in recent reports\,\cite{Glazov2019, Perea-Causin2019}.

\begin{figure}[t]
	\centering
			\includegraphics[width=8.4 cm]{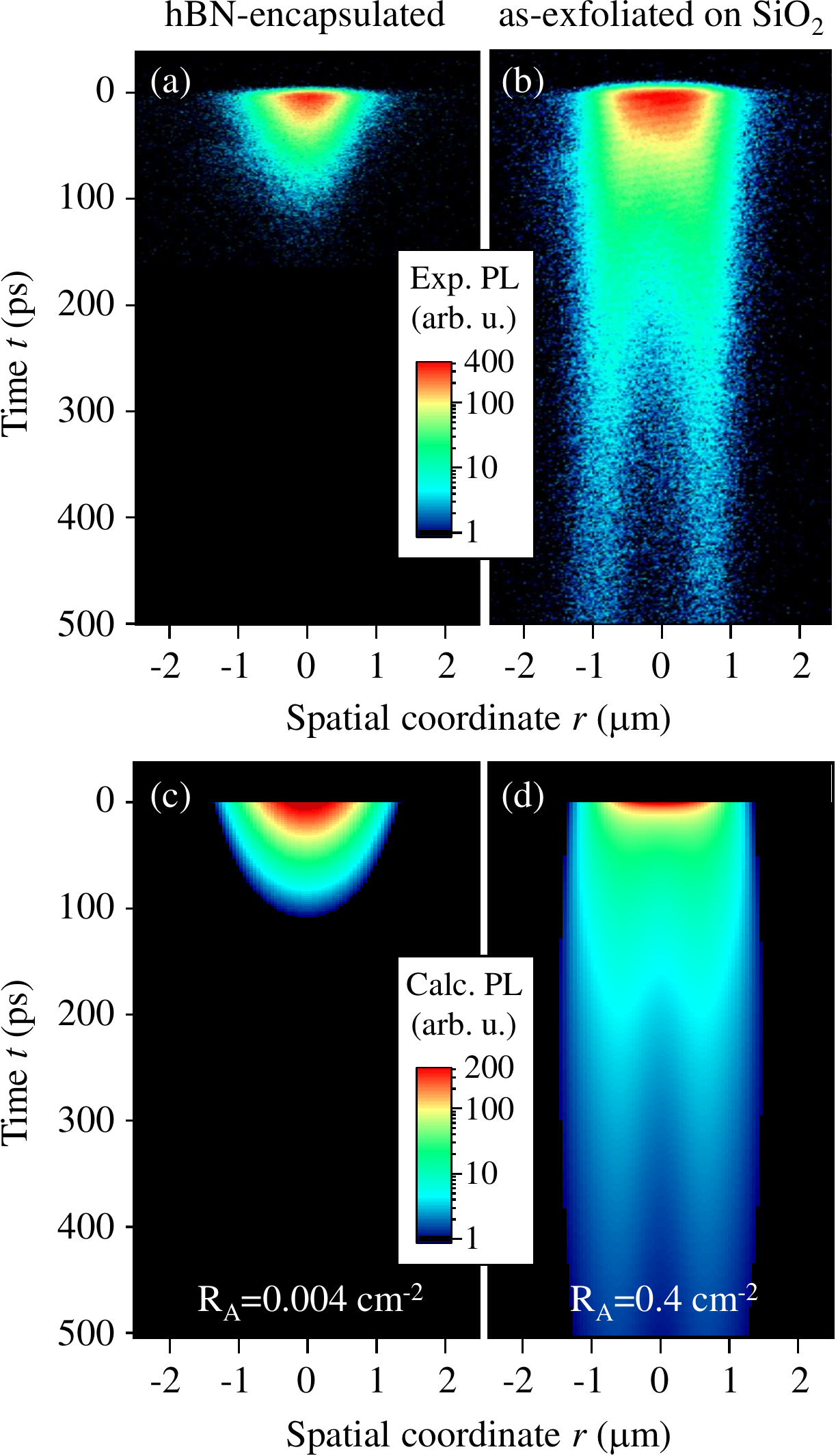}
		\caption{Spatially- and time-resolved images of the emitted luminescence at elevated excitation density of 0.6$\times$10$^{11}$\,cm$^{-2}$ electron-hole pairs per pulse for a hBN-encapsulated WS$_2$ monolayer (a) and an as-exfoliated sample on SiO$_2$/Si substrate (b). 
		The data are shown on the same spatial and temporal axes for direct comparison. 
		The emission intensity is presented in false-color on logarithmic scale in arbitrary units; the scale is multiplied by factor of 4 in panel (a).
		Corresponding results of the theoretical calculations using the model and key parameters from Ref.\,\onlinecite{Perea-Causin2019} are presented in (c) and (d), with the experimentally determined Auger coefficients of 0.004 and 0.4\,cm$^2$/s, respectively.
		}
	\label{fig7}
\end{figure} 

Notably, significant non-equilibrium populations of hot phonons and excitons can lead to the appearance of ring-like halo shapes in the sample emission with strongly suppressed luminescence in the center of the spot. 
This phenomenon was previously observed in as-exfoliated WS$_2$ and WSe$_2$ monolayers exhibiting efficient Auger recombination\,\cite{Kulig2018}.
As presented further above and illustrated in Fig.\,\ref{fig6}, Auger scattering is strongly suppressed in hBN-encapsulated samples by about two orders of magnitude.
Further considering the efficient diffusion counteracting exciton drift currents and shorter lifetime effectively limiting the observation window in hBN-encapsulated WS$_2$ monolayers, one may thus not expect a rapid evolution of the exciton distribution into halos.

A direct comparison of spatially- and time-resolved streak images from hBN-encapsulated and as-exfoliated WS$_2$ samples is presented in Figs.\,\ref{fig7}\,(a) and (b), respectively.
Both images were recorded at the same estimated electron-hole pair density, taking into account differences in the effective absorption of the two structures at the pump laser energy.
For the as-exfoliated sample, the detected luminescence cross-section clearly illustrates the evolution of a Gaussian profile into a double peak structure, that would correspond to a ring in a two-dimensional distribution.
In stark contrast to that, the emission of the hBN-encapsulated sample retains the initial shape of a peak during decay and does not exhibit hallmarks of a halo formation.

For a quantitative support of the proposed interpretation of our findings we employ the theoretical model outlined in Ref.\,\onlinecite{Perea-Causin2019} for the two studied cases, fixing the Auger coefficients and exciton lifetimes to experimentally determined values to directly illustrate the associated consequences. 
This microscopic model is similar to the Bloch-equation approach briefly outlined in Section\,\ref{linear-A} and described in more detail in the corresponding references\,\cite{Perea-Causin2019,Rosati2019}.
The method is further extended to account for Auger recombination and the subsequent emission of hot optical phonons, which are reabsorbed by the excitons, heating them up and leading to the formation of halos in the exciton spatial profile. 
For the discussion of the additional phonon-induced contributions to the exciton currents that can also lead to the formation of halos, please refer to the Ref.\,\onlinecite{Glazov2019} (also see Ref.\,\onlinecite{Kopelevich1996} for phonon-assisted processes in exciton transport in bulk semiconductors).
Here, these contributions are omitted due to slow phonon velocities compared to the average velocity of excitons\,\cite{Jin2014}.

The microscopic processes result in an unconventional diffusion that can be described with a modified Fick's law for the exciton current density $\textbf{j}(\textbf{r},t)$ as function of density $n_X(\textbf{r},t)$ and exciton temperature $T_X(\textbf{r},t)$, that reads:
\begin{equation}
\label{Xcurrent}
\textbf{j}(\textbf{r},t)=-D\nabla_{\textbf{r}}n_X(\textbf{r},t)-\sigma s\nabla_{\textbf{r}}T_X(\textbf{r},t),
\end{equation}
with the diffusion coefficient $D$,  the effective conductivity $\sigma$ and the Seebeck coefficient $s$ that can be calculated microscopically taking into account exciton-phonon scattering.
Note that the coefficients $D$ and $\sigma s$ implicitly depend on $n_X$ and $T_X$, respectively, as shown in detail in Ref.\,\onlinecite{Perea-Causin2019}.
The first term of Eq.\,\ref{Xcurrent} corresponds to the conventional diffusion law and the second one appears for a non-uniform exciton temperature distribution $T_X(\textbf{r},t)$ driving the excitons from ``hotter'' towards ``colder'' regions.
The local increase of the exciton temperature is caused by the reabsorption of hot optical phonons that are emitted by high-energy Auger-scattered excitons. We trace these complex processes by numerically solving the spatio-temporal Bloch-equations described in Ref.\,\onlinecite{Perea-Causin2019}.

The results of the calculations are presented in Figs.\,\ref{fig7}\,(c) and (d) for two different Auger coefficients of 0.004 and 0.4\,cm$^2$/s, respectively.
While the PL decay in as-exfoliated samples is already well described by Auger recombination due to much slower low-density lifetime, a decay time of 20\,ps is included in the model for hBN-encapsulated samples to account for a more efficient exciton trapping, as discussed in Section\,\ref{linear-C}.
Overall, the model accurately reproduces the main features observed in the experiment.
In particular, the experimentally determined decrease of the Auger coefficient by two orders of magnitude is found to be sufficient to completely suppress halo formation for the case corresponding to hBN-encapsulated sample, confirming our main interpretation.
We note that while halo formation may still appear at even higher densities, the exciton ionization at the Mott transition at about 10$^{13}$\,cm$^{-2}$ estimated from Ref.\,\onlinecite{Steinhoff2017} should provide a natural upper limit.
Overall, we conclude that all non-linear phenomena in the exciton propagation driven by Auger recombination are found to be strongly suppressed upon hBN-encapsulation.

\section{Conclusions}
\label{conclusions}

In this work we have explored linear and non-linear diffusion of excitons in hBN-encapsulated WS$_2$ monolayer materials at room temperature and ambient conditions.
The encapsulation suppresses dielectric disorder that is otherwise inherent for typical as-exfoliated samples strongly affecting propagation of the photoexcitations\,\cite{Raja2019}.
Exciton diffusion was experimentally studied by the means of spatially- and time-resolved photoluminescence imaging microscopy,  varying excitation densities across many orders of magnitude.
The observations were analyzed using a combination of numerical and analytical theoretical approaches providing insight into the underlying microscopic mechanisms.
In summary, we arrive at the following main conclusions:

\begin{enumerate}[label=\textbf{\roman*.}]
	\item \textbf{Exciton diffusion is highly efficient} in the linear regime with diffusion coefficients on the order of 5 to 10\,cm$^2$/s, that correspond to an effective mean free path of about 20\,nm at room temperature (Sec.\,\ref{results}).
	For an accurate description of the exciton propagation we demonstrate the influence of the complex exciton bandstructure of WS$_2$ and the particular importance of spin- and momentum-dark states (Sec.\,\ref{linear-A}). 
	\item \textbf{Free electron-hole plasma can facilitate rapid diffusion} at sufficiently low excitation densities due to entropy ionization of excitons in hBN-encapsulated samples with binding energies on the order of 200\,meV.
	We show that already at injection densities below 10$^{10}$\,cm$^2$ we could expect sizable fractions of free charge carriers leading to faster effective diffusion of the total population (Sec.\,\ref{linear-B}).
	\item \textbf{Short population lifetime accompanies increased diffusion} upon encapsulation, as we demonstrate through consistent observations based on a large number of individual measurements.
	We provide a pathway to understand these findings by illustrating a possible relationship between rapid propagation and fast capture into non-radiative trap regions (Sec.\,\ref{linear-C}).
	\item \textbf{Propagation phenomena driven by Auger-recombination are strongly suppressed} in contrast to as-exfoliated samples.
	While we observe a linear increase of the effective diffusion coefficient at elevated excitation densities due to Auger processes in analogy to the behavior of as-exfoliated monolayers, the slope is lower by about two orders of magnitude (Sec.\,\ref{non-linear-A}).
	It is directly proportional to the strongly reduced efficiency of the Auger scattering that influences measured effective diffusion coefficients\,\cite{Warren2000,Kulig2018}.
	As a consequence, all phenomena associated with the excitonic Auger processes\,\cite{Glazov2019,Perea-Causin2019} are far less efficient or largely suppressed, including ring-shaped halo formation in the exciton emission (Sec.\,\ref{non-linear-B}).	
\end{enumerate}

Overall, we find rich and non-trivial exciton propagation dynamics in monolayer samples with suppressed disorder and present pathways towards microscopic understanding of the experimental observations.
Notably, the phenomena associated with complexities of the electronic and excitonic bandstructure, presence of both excitons and free charge carriers, as well as inter-excitonic interactions are shown to be of particular importance.
In conclusion, a better understanding of the inherent properties of van der Waals monolayer semiconductors should strongly motivate fundamental investigation, manipulation, and control of interacting quasiparticles in ultra-thin materials, stimulating future research.

\section{Acknowledgments}
We thank our colleagues Archana Raja, Sivan Refaely-Abramson, and Malte Selig for helpful discussions.
Financial support by the DFG via Emmy Noether Grant CH 1672/1-1 and Collaborative Research Center SFB 1277 (B05) is gratefully acknowledged. 
The project has also received funding from the European Union`s Horizon 2020 research and innovation program under grant agreement No 785219 (Graphene Flagship) and the Swedish Research Council (VR, project number 2018–00734).
Growth of hexagonal boron nitride crystals was supported by the
Elemental Strategy Initiative conducted by the MEXT, Japan and the CREST
(JPMJCR15F3), JST.
Theoretical work of M.M.G. was partially supported by Russian Science Foundation (Project No. 19-12-00051).

%

~\\
\newpage

\appendix
\hypertarget{AppA}{\section{Exciton entropy ionization}}

Here we present a more detailed discussion of the mass action law.
It is followed by the application for a multi-band scenario in thermodynamic equilibrium including a full set of equations for the studied case of WS$_2$ monolayer and their numerical solutions. 
Finally we show that an arbitrary multi-band scenario can be reduced to an effective two-component model when only the total electron, hole, and exciton densities are considered.

\subsection {Derivation of the mass action law}

The mass action law was originally proposed in the XIX century to describe ratios of chemical reaction components in the steady state.
It was subsequently adopted for the description of exciton quasiparticles in semiconductors formed from free electrons and holes\,\cite{Mock1978,Kaindl2009}.
Here we recall the derivation of the mass-action law to motivate its subsequent expansion for a more complex multi-valley scenario used to describe the studied case of hBN-encapsulated WS$_2$ monolayers.

First, we consider the free energy density of a two-band system consisting of free electrons, free holes, and excitons with the respective densities $n_e$, $n_h$, and $n_X$. 
For elevated temperatures and moderate to low densities ($k_B T \gg n/\mathcal D$, see below) we can use the classical gas approximation for exciton and free charge carrier statistics.
For the typical parameters of TMDCs monolayers it should be valid up to the densities of several 10$^{12}$\,cm$^{-2}$ at room temperature.
The free energy then reads:
\begin{align}
\label{AppFreeE}
\mathcal F&=n_e k_B T \left(\ln\frac{n_e}{\mathcal D_e k_B T}-1\right) \notag \\ 
&+\,n_h k_B T \left(\ln\frac{n_h}{\mathcal D_h k_B T}-1\right) \notag \\ 
&+\,n_X k_B T \left(\ln\frac{n_X}{\mathcal D_X k_B T}-1\right)-E_b n_X.
\end{align}
Here, T is the carrier temperature, $\mathcal D_{e,h,X}$ are the respective densities of states (in two dimensions equal to $gm/2\pi\hbar^2$ with center-of-mass $m$ and degeneracy $g$), and $E_b>0$ is the exciton binding energy. 
In order to determine the densities of the particles we need to minimize the free energy under the following conditions:
\begin{align}
N_{tot} = n_X + n_h,\\
n_e = n_h + n_{e}^{(0)}.
\end{align}
The first condition means that the total density $N_{tot}$ of optically excited electron-hole pairs is redistributed between the excitons and plasma.
The second condition implies that the total density of electrons in the system is equal to the density of photogenerated electrons (given by
the density of free holes) plus the resident electrons $n_{e}^{(0)}$ provided by doping.
Introducing the Lagrange multipliers $\lambda_1$ and $\lambda_2$ and minimizing
\begin{align}
L(n_e, n_h, n_X;\lambda_1,\lambda_2)=\mathcal F+\lambda_1(N_{tot}-n_X-n_h)\notag  \\+ \lambda_2(n_e-n_h-n_{e}^{(0)})
\end{align}
by using the conditions $\partial L/\partial n_e =0$, $\partial L/\partial n_h =0$, and $\partial L/\partial n_X =0$ we obtain the following set of equations:
\begin{subequations}
\begin{align}
\label{L1}
0&=k_B T \left(\ln\frac{n_e}{\mathcal D_e k_B T}\right)+\lambda_2 \\
\label{L2}
0&=k_B T \left(\ln\frac{n_h}{\mathcal D_h k_B T}\right)-\lambda_1-\lambda_2 \\
\label{L3}
0&=k_B T \left(\ln\frac{n_X}{\mathcal D_X k_B T}\right)-E_B-\lambda_1 
\end{align}
\end{subequations}
Excluding $\lambda_1$ and $\lambda_1$ from the Eqs.\,(\ref{L1})-(\ref{L3}) we obtain the mass action law, also known as the Saha-equation:
\begin{align}
\label{AppSaha}
\frac{n_e n_h}{n_X}=\frac{\mathcal D_e \mathcal D_h}{\mathcal D_X}k_B T e^{-E_b/k_B T}\equiv S.
\end{align}
Here, one can use the parameter $S$ for the right hand side of Eq.\,(\ref{AppSaha}), that becomes a constant in units of density for a given set of material parameters and fixed temperature.
Introducing the exciton fraction $\alpha_X=n_X/(n_X+n_h)$, i.e., the ratio of exciton density to the total number of photoexcited electron-hole pairs (excluding residual electrons from doping), we obtain the analytical solution of the Eq.\,(\ref{AppSaha}):
\begin{align}
\label{AppSahaSol}
\alpha_X=1+\frac{S+n_{e}^{(0)}}{2N_{tot}}-\sqrt{\left(\frac{S+n_{e}^{(0)}}{2N_{tot}}\right)^2+\frac{S}{N_{tot}}}.
\end{align}

\begin{figure}[h]
	\centering
			\includegraphics[width=8.0 cm]{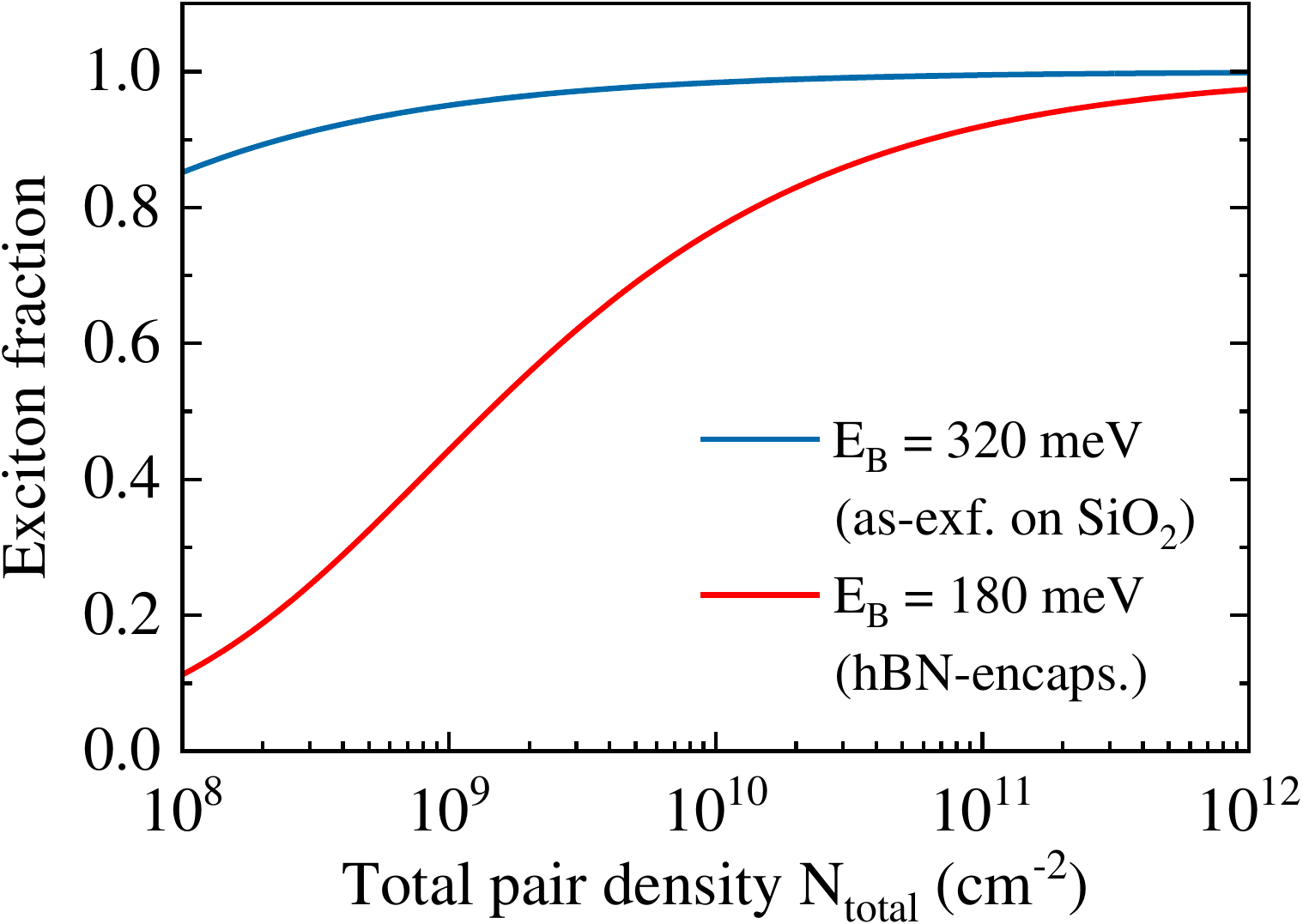}
		\caption{Exciton fractions calculated from the Saha-equation for two different values of exciton binding energies that are typical for WS$_2$ monolayer samples.}
	\label{figA0}
\end{figure}
Exemplary cases are illustrated in Fig.\,\ref{figA0} using parameters typical for bright $K-K$ excitons in hBN-encapsulated and SiO$_2$-supported, as-exfoliated WS$_2$ monolayers\,\cite{Goryca2019, Wang2018}.
Density-dependent exciton fractions are obtained from Eqs.\,\eqref{AppSaha} and \eqref{AppSahaSol} by setting the reduced exciton mass to 0.18\,$m_0$, the temperature to 290\,K, and the exciton binding energies either to 180\,meV or 320\,meV.
The corresponding Saha parameters $S$ are 7\,$\times$\,10$^{8}$\,cm$^{-2}$ and 2.6\,$\times$\,10$^{6}$\,cm$^{-2}$, respectively.
 
\subsection {Multi-band 2D semiconductor}

\begin{figure*}[ht]
	\centering
			\includegraphics[width=16.8 cm]{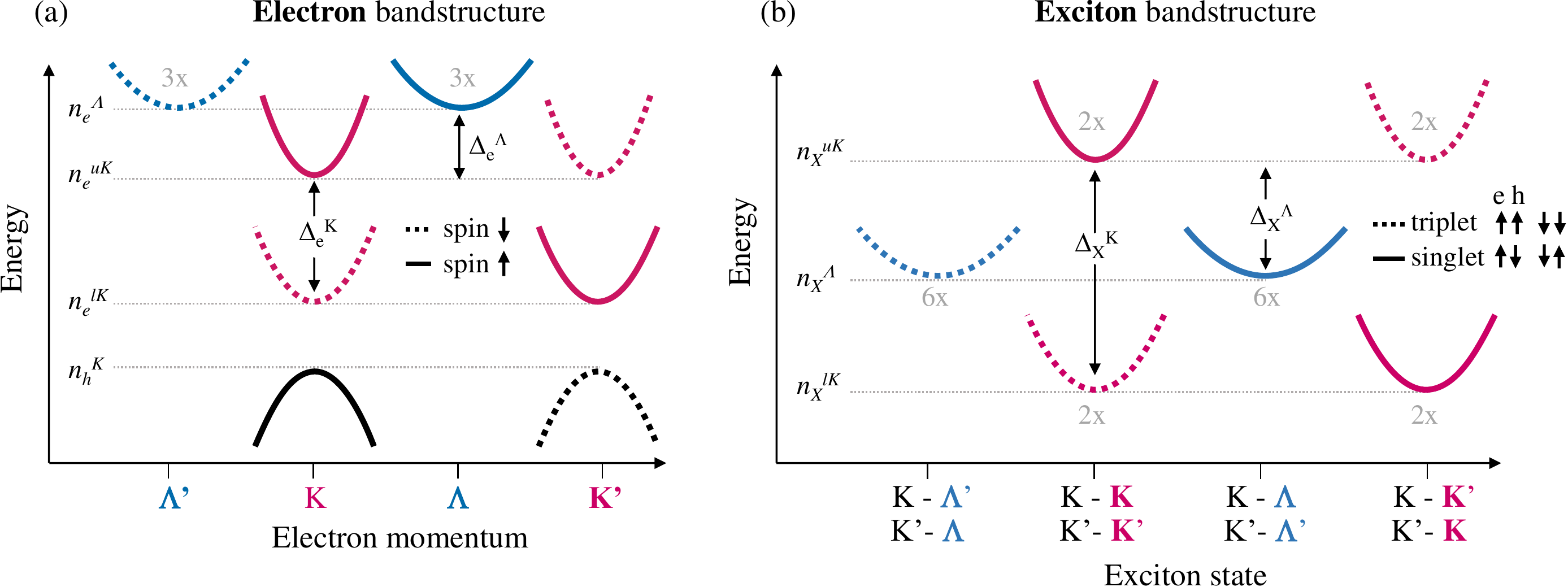}
		\caption{Schematic representations of the electron (a) and exciton (b) bandstructures of hBN-encapsulated WS$_2$ monolayer.
		The excitons are labeled according to the respective electronic transitions from $K$ or $K'$ in the valence band to $K$, $K'$, $\Lambda$, or $\Lambda'$ in the conduction band.
		Corresponding parameters of the relevant states are given in the tables \ref{tabA1} and \ref{tabA2}.
		}
	\label{figA1}
\end{figure*}

We now consider a more complex multi-valley bandstructure of the studied WS$_2$ monolayers. 
Following the previous section we include all relevant electron, hole, and exciton states in the free energy density in analogy to Eq.\,(\ref{AppFreeE}).
The single-particle electron bandstructure is schematically illustrated in Fig.\,\ref{figA1}\,(a).
For the electrons the bands are: upper and lower spin-split bands at $K$ and $K'$ (with densities $n_{e}^{uK}$ and $n_{e}^{lK}$) as well as three-fold-degenerate $\Lambda$ and $\Lambda'$ valleys (summarized by $n_e^{\Lambda}$).
For the holes we include only the upper valleys at $K$ and $K'$ ($n_{h}^{K}$) due to a very large spin-splitting in the valence band. 

The exciton bandstructure is schematically presented in Fig.\,\ref{figA1}\,(b).
The exciton states are composed from combinations of the holes at $K$ and $K'$ with electrons in upper and lower $K$ and $K'$ as well as in $\Lambda$ and $\Lambda'$ valleys (see Fig.\,\ref{fig3}\,(a) in the main text).
They are denoted by the respective electronic transitions and grouped according to their masses and relative energies.
The terms ``singlet'' and ``triplet'' correspond to the exciton total spin from the combined electron and hole constituents.
The exciton states are: $K-K$, $K'-K'$ singlet and $K-K'$, $K'-K$ triplet excitons with the electron in the \textit{upper} $K$ and $K'$ conduction bands ($n_{X}^{uK}$), $K-K$, $K'-K'$ triplet and $K-K'$, $K'-K$ singlet excitons with the electron in the \textit{lower} $K$ and $K'$ conduction bands ($n_{X}^{lK}$), as well as three-fold-degenerate $K-\Lambda$, $K'-\Lambda'$  singlets and $K-\Lambda'$, $K'-\Lambda$ triplets (summarized by $n_{X}^{\Lambda}$) with the electrons in the $\Lambda$ and $\Lambda'$ valleys.

The relevant parameters of the electron, hole, and exciton states are summarized in Tables \,\ref{tabA1} and \,\ref{tabA2}.
Included are total masses $m$ (in units of free electron mass $m_0$), energy offsets $\Delta$ with respect to the upper state at $K$, binding energies for the excitons $E_b$ and the degeneracy factors.
The values for the single particle conduction and valence band states are taken from ab-initio literature results\,\cite{Kormanyos2015}.
The exciton binding energies are calculated in the effective mass approximation using two-dimensional model for the Coulomb interaction\,\cite{Rytova1967,Keldysh1979} with hBN as a dielectric screening medium.
We note that while the resulting electron and exciton bandstructures should be realistic and representative, the specific values for the conduction band structure of the TMDCs are still under discussion in the community, both from theoretical and experimental perspectives.
Thus, one could generally expect deviations from the numbers given in Tables \ref{tabA1} and \ref{tabA2}.
Nevertheless, as we also discuss in the main text, the key conclusions appear to be robust with respect to reasonable variations of these parameters.

The resulting expression for the free energy density then reads:
\begin{align}
\label{AppFullFreeE}
\mathcal F&=n_{e}^{uK} k_B T \left(\ln\frac{n_{e}^{uK}}{\mathcal D_{e}^{uK} k_B T}-1\right) \\ 
&+\,n_{e}^{lK} k_B T \left(\ln\frac{n_{e}^{lK}}{\mathcal D_{e}^{lK} k_B T}-1\right)+n_{e}^{lK}\Delta_{e}^{K} \notag \\ 
&+\,n_{e}^{\Lambda} k_B T \left(\ln\frac{n_{e}^{\Lambda}}{\mathcal D_{e}^{\Lambda} k_B T}-1\right)+n_{e}^{\Lambda}\Delta_{e}^{\Lambda} \notag \\ 
&+n_{h}^{K} k_B T \left(\ln\frac{n_{h}^{K}}{\mathcal D_{h}^{K} k_B T}-1\right) \notag \\ 
&+n_{X}^{uK} k_B T \left(\ln\frac{n_{X}^{uK}}{\mathcal D_{X}^{uK} k_B T}-1\right)-n_{X}^{uK}E_{b}^{uK} \notag \\ 
&+n_{X}^{lK} k_B T \left(\ln\frac{n_{X}^{lK}}{\mathcal D_{X}^{lK} k_B T}-1\right)+n_{X}^{lK}(-E_{b}^{lK}+\Delta_{e}^{K}) \notag \\ 
&+n_{X}^{\Lambda} k_B T \left(\ln\frac{n_{X}^{\Lambda}}{\mathcal D_{X}^{\Lambda} k_B T}-1\right)+n_{X}^{\Lambda}(-E_{b}^{\Lambda}+\Delta_{e}^{\Lambda}) \notag
\end{align}

Here, $E_b$ and $\Delta_e$ with corresponding superscripts are the exciton binding energies, and \textit{conduction band} energy offsets relative to the upper $K$ valley state, respectively.
Note that for the exciton contributions, one can alternatively also use the \textit{exciton} energy offsets $\Delta_X$ relative to the $K-K$ singlet state minus the exciton binding energy of the latter ($E_{b}^{uK}$), c.f. Fig.\,\ref{figA1}\,(b).
The densities of states, denoted by $\mathcal D$, depend on the values of the total mass $m$ and are proportional to the degeneracy of the corresponding state.
They are given by:
\begin{equation}
\mathcal D_{e}^{uK}=2\frac{m_{e}^{uK}}{2\pi\hbar^2},~~\mathcal D_{e}^{lK}=2\frac{m_{e}^{lK}}{2\pi\hbar^2},~~\mathcal D_{e}^{\Lambda}=6\frac{m_{e}^{\Lambda}}{2\pi\hbar^2},\notag
\end{equation}
\begin{equation}
\label{AppFullDOS}
\mathcal D_{h}^{K}=2\frac{m_{h}^{K}}{2\pi\hbar^2}
\end{equation}
\begin{equation}
\mathcal D_{X}^{uK}=4\frac{m_{X}^{uK}}{2\pi\hbar^2},~~\mathcal D_{X}^{lK}=4\frac{m_{X}^{lK}}{2\pi\hbar^2},~~D_{X}^{\Lambda}=12\frac{m_{X}^{\Lambda}}{2\pi\hbar^2}.\notag
\end{equation}
The conservation of the total particle density $N_{tot}$ and charge neutrality condition read:
\begin{align}
\label{AppCond1}
N_{tot}=n_{h}^{K}+n_{X}^{uK}+n_{X}^{lK}+n_{X}^{\Lambda} \\
\label{AppCond2}
n_{e}^{uK} + n_{e}^{lK} + n_{e}^{\Lambda} + n_{h}^{(0)} = n_{h}^{K} + n_{e}^{(0)}
\end{align}
In Eq.\,(\ref{AppCond2}) one can include residual electron and hole doping densities $n_{e}^{(0)}$ and $n_{h}^{(0)}$, respectively.
In the following they are set to zero to illustrate the case of an undoped system.

\begin{table}[h]
\centering 
\begin{tabular}{l | c  c  c} 
\hline\hline\\
\textbf{Electron} state~~&~~$m/m_0$&~~$\Delta_e$(meV)&~~deg.\\ [0.5ex] 
\hline \\ 
CB, upper $K$, $K'$~~&0.27& 0 & 2x \\ [0.5ex] 
CB, lower $K$, $K'$~~&0.36& -31 & 2x \\ [0.5ex] 
CB, $\Lambda$, $\Lambda'$~~&0.64& 27 & 6x \\ [0.5ex] 
VB, upper $K$, $K'$~~&0.36& - & 2x \\ [1ex] 
\hline 
\end{tabular}
\caption{Electron bandstructure parameters for the electron states in conduction (CB) and upper valence (VB) bands.
}
\label{tabA1} 
\end{table}

In analogy to the derivation of the mass action law for the two-band scenario discussed in the previous section, we combine Eqs.\,(\ref{AppFullFreeE}),(\ref{AppCond1}), and (\ref{AppCond2}) using Lagrange formalism and obtain the following set of equations:
\begin{subequations}
\begin{align}
\label{AppFullEq1}
0&=k_B T \left(\ln\frac{n_{e}^{uK}}{\mathcal D_{e}^{uK} k_B T}\right)+\lambda_2\\
0&=k_B T \left(\ln\frac{n_{e}^{lK}}{\mathcal D_{e}^{lK} k_B T}\right)+\Delta_{e}^{K}+\lambda_2\\
0&=k_B T \left(\ln\frac{n_{e}^{\Lambda}}{\mathcal D_{e}^{\Lambda} k_B T}\right)+\Delta_{e}^{\Lambda}+\lambda_2\\
0&=k_B T \left(\ln\frac{n_{h}^{K}}{\mathcal D_{h}^{K} k_B T}\right)-\lambda_1-\lambda_2\\
0&=k_B T \left(\ln\frac{n_{X}^{uK}}{\mathcal D_{X}^{uK} k_B T}\right)-E_{b}^{uK}-\lambda_1\\
0&=k_B T \left(\ln\frac{n_{X}^{lK}}{\mathcal D_{X}^{lK} k_B T}\right)+\Delta_{e}^{K}-E_{b}^{lK}-\lambda_1\\
\label{AppFullEq7}
0&=k_B T \left(\ln\frac{n_{X}^{\Lambda}}{\mathcal D_{X}^{\Lambda} k_B T}\right)+\Delta_{e}^{\Lambda}-E_{b}^{\Lambda}-\lambda_1
\end{align}
\end{subequations}

\begin{table}[h]
\centering 
\begin{tabular}{l | c  c  c  c} 
\hline\hline\\
\textbf{Exciton} state~~&~~$m/m_0$&$E_b$(meV)&~~$\Delta_X$(meV)~~&~~deg.\\ [0.5ex] 
\hline \\ 
e: upper $K$, $K'$~~&0.63& 183 & 0 & 4x \\ [0.5ex] 
e: lower $K$, $K'$~~&0.72& 202 & -50 & 4x \\ [0.5ex] 
e: $\Lambda$, $\Lambda'$~~&1.0& 236 & -26 & 12x \\ [1ex] 
\hline 
\end{tabular}
\caption{Exciton bandstructure parameters for the excitons with the hole in the upper $K$ and $K'$, labeled according to the states of the electron (e) constituents.
}
\label{tabA2} 
\end{table}

\begin{figure}[h]
	\centering
			\includegraphics[width=8.4 cm]{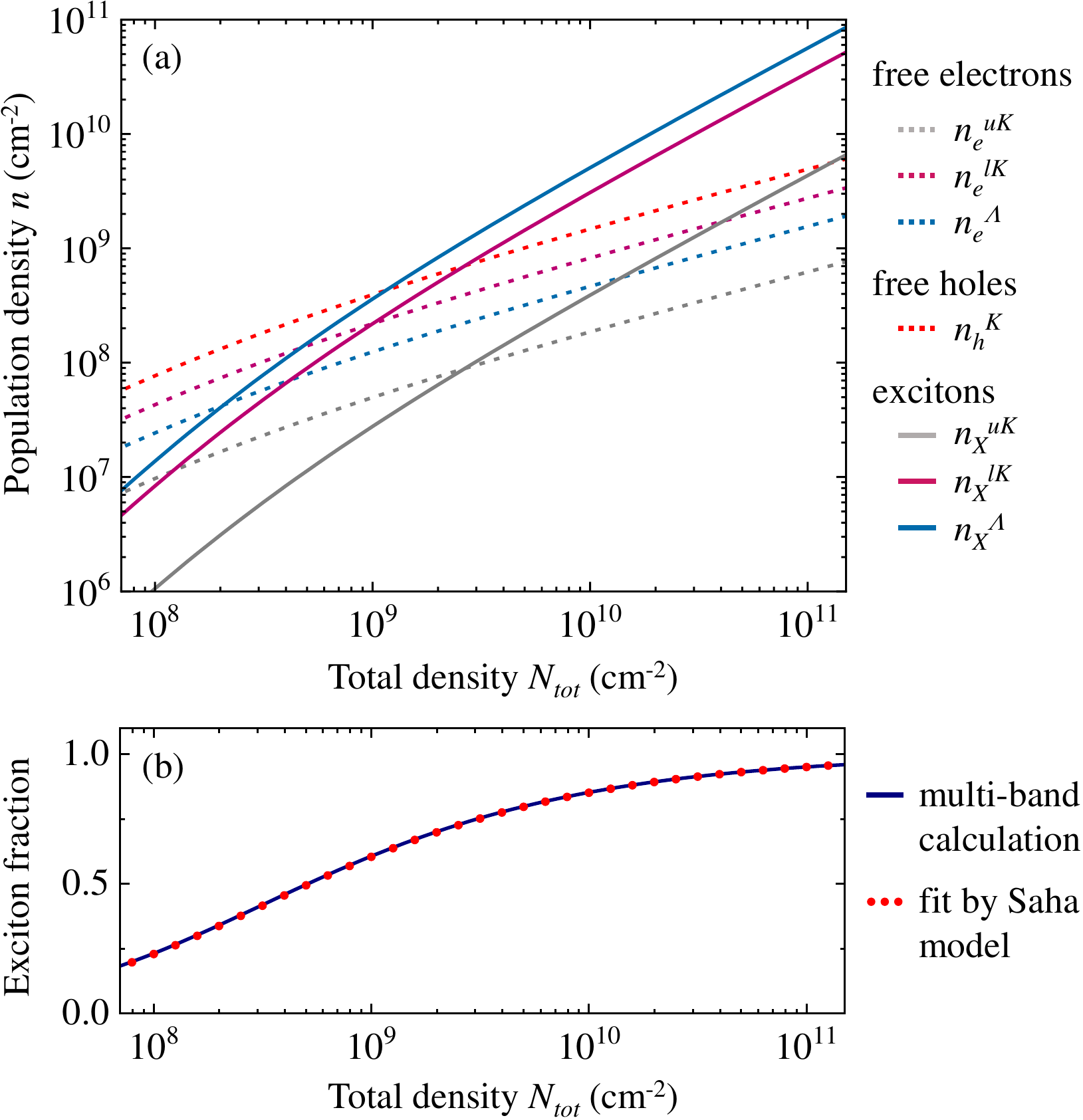}
		\caption{(a) Numerical results of the multi-band calculation of the individual equilibrium densities of free electrons, holes, and excitons with the bandstructure and exciton parameters of hBN-encapsulated WS$_2$ at $T$\,=\,300\,K.
		(b) Total exciton fraction from the multi-band calculations and the corresponding fit by the two-band Saha model (Eq.\,(\ref{AppSaha}) with $S$=2.5$\times$10$^8$\,cm$^{-2}$.
		}
	\label{figA2}
\end{figure}

The Eqs.\,(\ref{AppFullEq1})-(\ref{AppFullEq7}) are numerically solved for the carrier temperature of $T$\,=\,300\,K.
The resulting individual densities of electrons, holes, and excitons are presented in Fig.\,\ref{figA2}\,(a) for the range of total densities between 10$^{8}$ and 10$^{11}$\,cm$^{-2}$.
At elevated densities beyond 10$^{10}$\,cm$^{-2}$ the majority of the carriers are indeed bound to excitons, formed by the holes at $K$ and $K'$ and the electrons either at the lower spin-split $K$ and $K'$ or at $\Lambda$ and $\Lambda'$ valleys.
At lower densities relative populations of the free charge carriers increase.
In particular, free holes have the highest relative occupation for densities of 10$^{9}$\,cm$^{-2}$ and below.

The total exciton fraction can be evaluated by computing the sum of the $n_{X}^{uK}$, $n_{X}^{lK}$, and $n_{X}^{\Lambda}$ densities divided by the total density as shown in Fig.\,\ref{figA2}\,(b).
The obtained fraction can be then described by a simpler, analytical case of a two-band Saha model using Eq.\,(\ref{AppSaha}) and a Saha-parameter $S$ (here, $S$=2.5$\times$10$^8$\,cm$^{-2}$).
As illustrated in the main text, this can be advantageous for the subsequent implementation and estimation of the exciton fraction that we then use to analyze the combined diffusion dynamics of the exciton and free carrier mixtures.
Below we demonstrate that such a description is indeed possible for an arbitrary multi-band system.

\subsection {Saha parameter for an arbitrary multi-band scenario}
Following the above arguments we show that any multi-band scenario with an arbitrary number of electron ($n_{e,i}$), hole ($n_{h,j}$), and exciton ($n_{X,k}$) states can be reduced to a simple Saha equation for the \textit{total} densities.
The general form of the free energy density reads:

\begin{align}
\label{AppGenfreeE}
\mathcal F&=\sum_i{n_{e,i} k_BT\,\left(\ln\frac{n_{e,i}}{\mathcal D_{e,i} k_BT}-1\right)+n_{e,i}E_{e,i}} \\
&+\sum_j{n_{h,j} k_BT\,\left(\ln\frac{n_{h,j}}{\mathcal D_{h,j} k_BT}-1\right)+n_{h,j}E_{h,j}} \notag \\
&+\sum_k{n_{X,k} k_BT\,\left(\ln\frac{n_{X,k}}{\mathcal D_{X,k} k_BT}-1\right)+n_{X,k}E_{X,k}} \notag
\end{align}
where $\mathcal D$ and $E$ are the densities of states and energy differences of the corresponding states, respectively (the later include the binding energies for excitons with a negative sign).
The total particle number $N_{tot}$ and the charge neutrality conservation conditions are:
\begin{align}
\label{AppGenCond}
N_{tot}=\sum_j{n_{h,j}}+\sum_k{n_{X,k}} \\
\sum_i{n_{e,i}}=\sum_j{n_{h,j}}.
\end{align}
With the Lagrange multipliers $\lambda_1$ and $\lambda_2$ we thus obtain
\begin{align}
L(n_{e,i},\dots, n_{h,j},\dots, n_{X,k},\dots;\lambda_1,\lambda_2)=\\
\mathcal F+\lambda_1(N_{tot}-\sum_j{n_{h,j}}-\sum_k{n_{X,k}})\notag  \\
+\lambda_2(\sum_i{n_{e,i}}-\sum_j{n_{h,j}}). \notag
\end{align}
Minimizing $L$ using conditions $\partial L/\partial n_e =0$, $\partial L/\partial n_h =0$, and $\partial L/\partial n_X =0$ provides a set of equations with the same structure for each $i$, $j$, and $k$:
\begin{subequations}
\begin{align}
\label{GenL1}
0&=k_B T \left(\ln\frac{n_{e,i}}{\mathcal D_{e,i} k_B T}\right)+E_{e,i}+\lambda_2 \\
\label{GenL2}
0&=k_B T \left(\ln\frac{n_{h,j}}{\mathcal D_{h,j} k_B T}\right)+E_{h,j}-\lambda_1-\lambda_2 \\
\label{GenL3}
0&=k_B T \left(\ln\frac{n_{X,k}}{\mathcal D_{X,k} k_B T}\right)+E_{X,k}-\lambda_1 
\end{align}
\end{subequations}
In analogy to Eq.\,(\ref{AppSaha}) we thus obtain a Saha-type equilibrium equation for any combination of $i$, $j$, and $k$ in the form of:
\begin{align}
\label{AppGenSaha}
\frac{n_{e,i} n_{h,j}}{n_{X,k}}=S(i,j;k),
\end{align}
where we introduced the partial Saha-parameters:
\[
S(i,j;k) =\frac{\mathcal D_{e,i} \mathcal D_{h,j}}{\mathcal D_{X,k}}k_B T e^{(-E_{e,i}-E_{h,j}+E_{X,k})/k_B T}.
\]

Our ultimate goal is to present a compact expression for the ratio $\bar n_e\bar n_h/\bar n_X$. To that end one can readily obtain
\begin{equation}
\label{int:S:1}
\frac{\bar n_e \bar n_h}{n_{X,k}} = \sum_{ij} S(i,j;k).
\end{equation}
Next, we sum inverse of Eq.~\eqref{int:S:1} over excitonic states $k$:
\begin{equation}
\label{int:S:2}
\frac{\bar n_{X}}{\bar n_e \bar n_h} = \sum_k \left[\sum_{ij} S(i,j;k)\right]^{-1}.
\end{equation}
As a result, we arrive at an effective two-band Saha equation 
\begin{align}
\label{AppTotalSaha}
\frac{\bar{n}_{e} \bar{n}_{h}}{\bar{n}_{X}}=\bar{S}
\end{align} 
with
\begin{align}
\label{AppTotalSahaParam}
\bar{S}^{-1}=\sum_{k}\left(\sum_{i,j}{\frac{\mathcal D_{e,i} \mathcal D_{h,j}}{\mathcal D_{X,k}}k_B T e^{\frac{-E_{e,i}-E_{h,j}+E_{X,k}}{k_B T}}}\right)^{-1}.
\end{align}
Then the analytical solution from Eq.\,(\ref{AppSahaSol}) can be readily used to compute density- and temperature-dependent total exciton fractions for any arbitrary number of individual electron, hole, and exciton bands.

\section{Exciton recombination limited by diffusion}\label{AppB}

In this section of the Appendix we present the general case of the exciton capture into non-radiative trap regions and discuss two limiting cases with respect to ratio of the trap size and the exciton mean free path.
First, let us assume that the trap creates the potential field $V(r)<0$, i.e., an attractive potential for excitons. 
We then solve the set of the continuity equations for the particle density $n$ (omitting hereafter all subscripts $X$ and $eh$ because the consideration below is valid regardless of type of particles we consider):
\begin{equation}
\label{continuity}
\frac{\partial n}{\partial t} + \bm \nabla \cdot \mathbf j =0,
\end{equation}
where $\mathbf j$ is the exciton current density.
The drift equation for the $\mathbf j$ [cf. Eq.\,\eqref{Xcurrent} and Ref.~\onlinecite{Glazov2019}] takes the following form in the vicinity of a trap:
\begin{align}
\mathbf j &= j_r \mathbf e_r + j_\phi \mathbf e_\phi,
\end{align}
with
\begin{align}
\label{current}
j_r &= D \frac{dn}{dr} + \frac{\tau}{m} \frac{dV}{dr} n, \quad  j_\phi=0.
\end{align}
Here the subscripts $r$ and $\phi$ denote the radial and azimuthal components of the current density, respectively ($\mathbf e_r$ is the radial and $\mathbf e_\phi$ is the azimuthal unit vectors).
The momentum scattering time of the particles is denoted by $\tau$ and their translational mass is represented by $m$.
Due to the continuity equation, the total radial component of the flux $\Phi=2\pi j_r$ does not depend on $r$ and is a constant. 
We can thus represent $\Phi=cn_0$, where $c$ is the capture rate to the trap and $n_0$ is the average exciton density far from the trap. 
Making use of the relation $D=k_B T \tau/m$, Eq.~\eqref{current} can be readily solved as
\begin{equation}
\label{density:distribution}
n(r) = n_0 \exp{\left(- \frac{V(r)}{k_B T}\right)} \left[1 - \frac{c}{2\pi D} \int_r^{\infty} \exp{\left( \frac{V(r)}{k_B T}\right)}\frac{dr}{r}\right].
\end{equation}

In order to calculate the capture rate $c$ and the non-radiative lifetime we need to impose the boundary condition at the trap. To that end we follow Ref.~\onlinecite{abakumov_perel_yassievich} and assume that the total flux is given by 
\begin{equation}
\label{beta:par}
\Phi = \beta n(r_0),
\end{equation}
where $\beta$ is the parameter of the trap which can be related to its geometrical capture rate, i.e., $\beta \sim \pi r_0 \langle v\rangle)$ with $\langle v\rangle$ representing the thermal velocity of the incoming particles. 
As a result we obtain
\begin{equation}
\label{c:gen}
\Phi=c n_0= \dfrac{\beta e^{-V(r_0)/k_B T}}{1+ \frac{\beta}{2\pi D}\int_{r_0}^{\infty} \exp{\left( \frac{V(r)-V(r_0)}{k_B T}\right)}\frac{dr}{r}}n_0.
\end{equation}

As discussed in the main manuscript we define the recombination time of the excitons in the traps according to:
\begin{equation}
\label{App-tau-1:1}
\frac{1}{\tau_r} \equiv \frac{N_{tr} \Phi}{n_0}
\end{equation}
with $N_{tr}$ being the density of the traps.
From the Eq.\,\eqref{c:gen} and the realistic limiting case $|V(r)| \ll k_B T$ for $r>r_0$ (i.e., between the traps) we find:
\begin{equation}
\label{tau-1:2}
\tau_r = \frac{1}{\beta N_{tr}} + \frac{1}{2\pi D N_{tr}} \ln{\left(\frac{1}{r_0\sqrt{N_{tr}}}\right)}.
\end{equation}
Here we cut-off the divergent integral at $r\sim 1/\sqrt{N_{tr}}$, corresponding to the mean distance between the traps.
The first term in Eq.~\eqref{tau-1:2} can be related to the geometric capture time by the trap 
\begin{equation}
\label{App-tau-cap}
\tau_{cap}  \sim \frac{1}{\pi r_0 \langle v \rangle N_{tr}}.
\end{equation}
~\\
The second term describes the capture limited by the diffusion and agrees up to the logarithmic factor with the estimate in the main text from Eq~\eqref{tau-1:1}:
\begin{equation}
\label{App-tau-D}
\tau_D = \frac{1}{2\pi D_X N_{tr}}.
\end{equation}
As one would also expect from qualitative arguments, the capture time is thus controlled by the longest process that is either the capture time $\tau_{cap}$ directly at the trap or the particle diffusion time $\tau_D$ to the trap. 
Then the ratio of the capture times is roughly given by the ratio of the mean free path $l$ to the trap radius $r_0$:
\begin{equation}
\label{ratio}
\frac{\tau_{cap}}{\tau_D} \sim \frac{\ell}{r_0}.
\end{equation}
Thus, for comparatively small traps with $r_0 \ll \ell$, the capture time at the trap $\tau_{cap}$ limits the lifetime and the non-radiative recombination rate thus does not depend on diffusion.
In contrast to that, for sufficiently large traps $r_0 \gg \ell$, the capture is limited by diffusion, as discussed in the main text.

\section{Long-term effects in hBN-encapsulated WS$_2$ samples}
\label{AppC}

\begin{figure}[ht]
	\centering
			\includegraphics[width=6 cm]{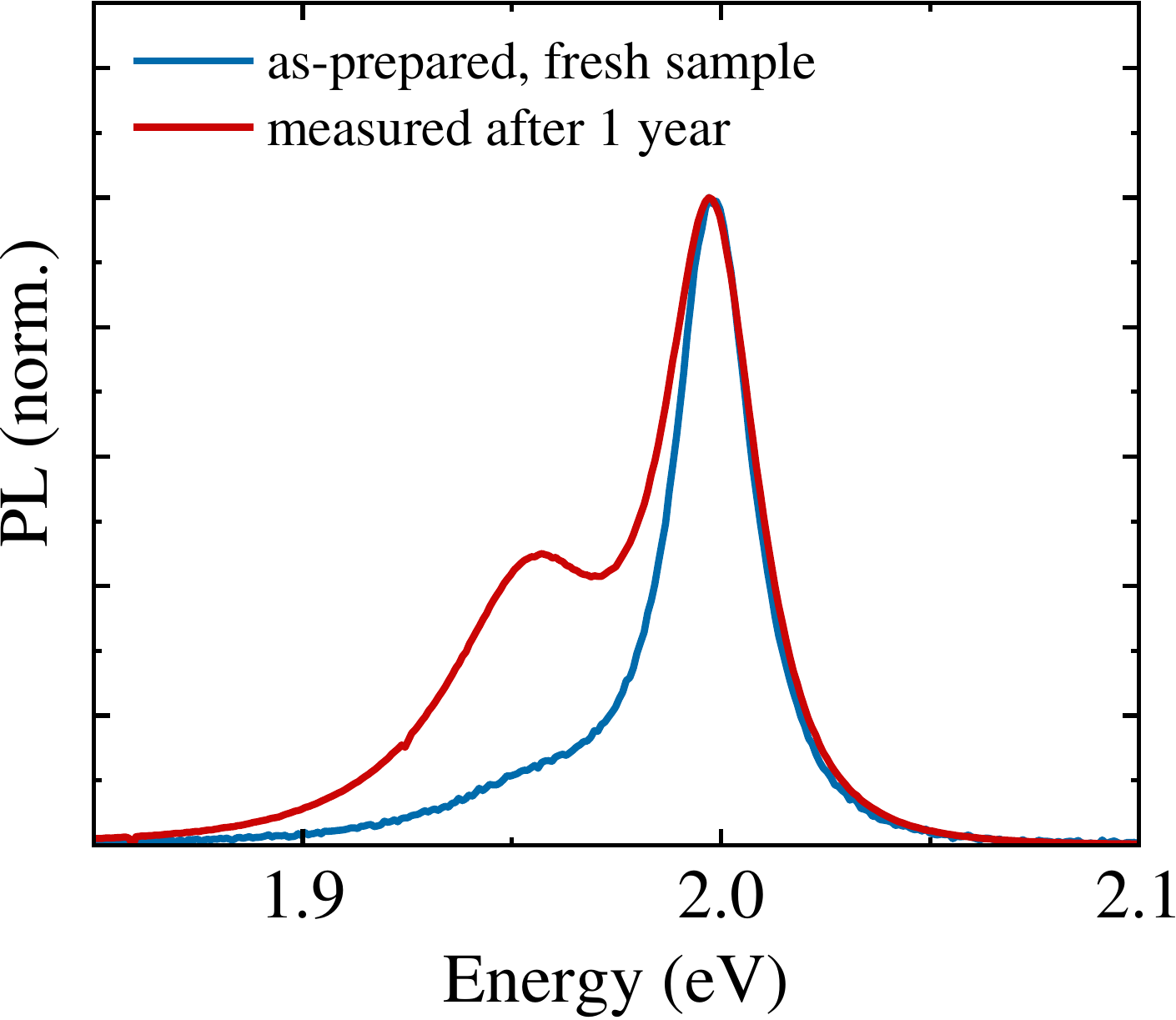}
		\caption{Luminescence spectra of a hBN-encapsulated WS$_2$ sample measured after the preparation and one year later for continuous-wave illumination in the linear regime.}
	\label{figA3}
\end{figure}
All measurements reported in this study were performed on freshly-prepared hBN-encapsulated WS$_2$ samples exhibiting dominant emission from the neutral exciton resonance. 
Measured about one year after the preparation the samples showed aging effects resulting in a more pronounced low-energy shoulder in the PL, as illustrated in Fig.\,\ref{figA3}, that is likely to stem from additional doping that increased over time.

\end{document}